\documentclass[11pt]{article}
\usepackage{jheppub}

\usepackage{epsfig}
\usepackage{amssymb}
\usepackage{amsmath}
\usepackage{amsfonts}

\usepackage{graphicx}
\usepackage[labelfont=bf]{caption}
\usepackage{subcaption}

\renewcommand{\d}{{\rm d}}
\newcommand{\ep}{\epsilon}
\newcommand{\Cf}{C_F}
\newcommand{\Ca}{C_A}

\newcommand{\mh}{M_H}
\newcommand{\mc}{m_c}
\newcommand{\Emax}{{E_{\rm max}}}
\newcommand{\fb}{{\rm fb}}
\newcommand{\gev}{{\rm GeV}}
\newcommand{\tev}{{\rm TeV}}
\newcommand{\ptj}{p_{t,j}}
\newcommand{\etaj}{\eta_{j}}

\newcommand{\MSbar}{ {\overline {\rm MS}} }
\newcommand{\scr}[2]{{p_{#1} \!\cdot\! p_{#2}}}
\newcommand{\FLM}{F_{\rm LM}}
\newcommand{\amp}{\mathcal{M}}
\newcommand{\cc}{{c\bar c}}
\newcommand{\ggh}{{ggH}}
\newcommand{\yuk}{{\rm Yuk}}
\newcommand{\interf}{{\rm Int}}
\newcommand{\pth}{p_{t,H}}

\newcommand{\LiTwo}{{{\rm Li}_2}}

\title{On the interference of $ggH$ and $\cc H$ Higgs production mechanisms and
  the determination of charm Yukawa coupling at the LHC
}

\preprint{
  TTP21-003,
  P3H-21-010
}

\author[a,b]{Wojciech Bizo\'{n},}
\author[a]{Kirill Melnikov,}
\author[a]{J\'er\'emie Quarroz}

\affiliation[a]{
  Institute for Theoretical Particle Physics (TTP),
  Karslruhe Institute of Technology,
  D-76128 Karlsrue, Germany
}

\affiliation[b]{
  Institute for Astroparticle Physics,
  Karslruhe Institute of Technology,
  D-76344 Eggenstein-Leopoldshafen, Germany
}

\abstract{ Higgs boson production in association with a charm-quark jet
  proceeds through two different mechanisms -- one that involves the
  charm Yukawa coupling and the other  that involves  direct Higgs
  coupling to gluons. The interference  of the two
  contributions requires a helicity flip and, therefore, cannot be
  computed with  massless charm quarks.  In this paper, we consider QCD
  corrections to the interference contribution starting from
  charm-gluon collisions with massive charm quarks  and taking the massless limit,
  $m_c \to 0$.
  The   behavior of QCD cross sections  in that limit differs from expectations
  based on the canonical QCD factorization. This implies that
  QCD corrections to the interference term necessarily involve
  logarithms of the ratio $\mh/m_c$ whose resummation is currently
  unknown.
  Although  the explicit next-to-leading order  QCD computation does confirm the presence of up to two powers of
  $\ln(\mh/m_c)$ in the interference contribution, their overall impact on the magnitude
  of  QCD corrections to the interference turns out to be moderate due to a cancellation between double and single logarithmic
  terms.
}

\begin{document}

\hfill{\today}
\maketitle

\clearpage

\section{Introduction}
\label{sec:intro}

Studies of Yukawa couplings play an important role in the verification
of the  mechanism of electroweak symmetry breaking as described by the
Standard Model.  By now, Higgs couplings to bottom and top quarks,
as well as to tau leptons and muons, have been measured
to a precision of about twenty percent \cite{Aaboud:2018zhk,Sirunyan:2018kst,Aaboud:2018pen,Sirunyan:2018cpi,Aad:2020xfq,Sirunyan:2018hbu}.  Within the error bars, the
measured values for all four Yukawa couplings are consistent with the
Standard Model predictions.

However, the Yukawa couplings to lighter fermions have not been studied
experimentally.  Although  it is generally agreed that the Yukawa
couplings of  electrons and up, down and strange  quarks can be observed if and
only if they enormously deviate from their  Standard Model values, the
situation with the charm Yukawa  is not so hopeless. In fact, it appears that
with the full LHC luminosity,  the charm Yukawa coupling can be measured if
its value  deviates from the Standard Model expectation  by an order one
factor \cite{Perez:2015lra}.   Different observables to measure the
charm Yukawa coupling at the LHC have been proposed; they include
inclusive  ($H \to c \bar c$)  and exclusive  ($H \to J/\psi + \gamma$ and similar)
decays of the Higgs boson
 \cite{kagan,Modak:2014ywa,Koenig:2015pha},  the modifications of the Higgs transverse momentum
 distribution \cite{Bishara:2016jga}  in the $gg \to H+X$ process and, finally,
Higgs boson production cross section in
association with a charm jet \cite{Brivio:2015fxa}.

\begin{figure}
  \centering
  \begin{subfigure}{0.66\textwidth}
    \centering
    \includegraphics[height=50pt]{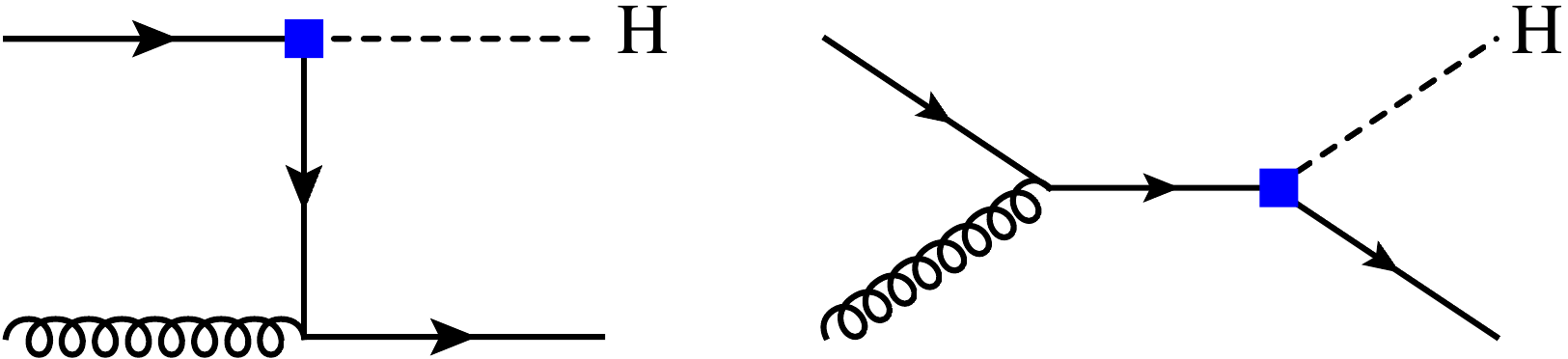}
    \caption{Yukawa coupling}
    \label{fig:feyn-diags-yuk}
  \end{subfigure}
  \begin{subfigure}{0.33\textwidth}
    \centering
    \includegraphics[height=50pt]{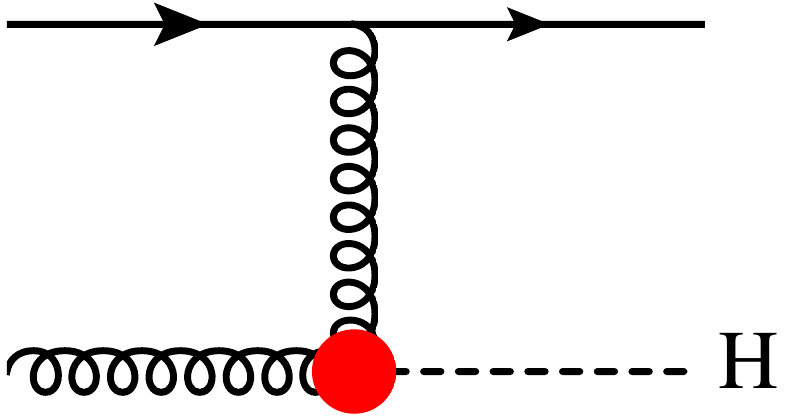}
    \caption{Effective $\ggh$ vertex}
    \label{fig:feyn-diags-ggh}
  \end{subfigure}
  \caption{Leading-order Feynman diagrams contributing to the
    $pp \to Hc$ process.       We distinguish two separate production mechanisms: one that is driven
    by the  Yukawa coupling
    (left) and the other one that requires direct coupling of Higgs to gluons
    (right).}
  \label{fig:feyn-diags}
\end{figure}

In this paper we focus on the latter process, $pp \to H+{\rm jet}_c$.
At leading order in
perturbative QCD, Higgs bosons are produced in association with
charm jets  in the partonic process  $cg \to Hc$. The amplitude of this process receives contributions
proportional to  the charm  Yukawa coupling and  to an
effective $ggH$ coupling
\begin{align}
  {\cal M} \sim g_{\yuk} {\cal M}_1 + g_{ggH} {\cal M}_2,
  \label{eq1aaa}
\end{align}
see  Figure~\ref{fig:feyn-diags}.  As a result, the
$pp \to H+{\rm jet}_c$ cross section contains the interference term
\begin{align}
  \sigma_{Hc} \sim g_{\yuk}^2 \; \tilde \sigma_1 + g_{ggH}^2 \; \tilde \sigma_2 + g_{\yuk} g_{ggH} \; \tilde \sigma_{\interf}.
  \label{eq2aaa}
\end{align}
It can be expected that a reliable  description of Higgs boson production in  association with a charm jet
can be obtained by systematically computing
the different terms in Eq.~(\ref{eq2aaa}) to higher orders in perturbative QCD. In fact,
it is emphasized  in Ref.~\cite{Brivio:2015fxa} that the largest theoretical uncertainty in using $H+{\rm jet}_c$ production cross section
to constrain charm Yukawa coupling is related to perturbative QCD uncertainties so that it seems natural
  to  compute  higher order QCD corrections to $\sigma_{Hc}$ in Eq.~(\ref{eq2aaa}).

However, pursuing this  program  for the interference term
in Eq.~(\ref{eq2aaa})  is quite subtle as  we now discuss.
Indeed, perturbative computations in QCD are performed with massless incoming partons.  In case of the massless
charm quark that, however, has  non-vanishing Yukawa coupling to the Higgs boson,
the interference term in Eq.~(\ref{eq2aaa}) vanishes and we obtain
\begin{align}
\lim_{m_c \to 0} \sigma_{Hc}\sim g_{\yuk}^2 \tilde \sigma_1 + g_{ggH}^2 \tilde \sigma_2.
\end{align}
  This happens because the Yukawa interaction flips
  charm's  helicity but the gluon-charm interaction  conserves it; hence, the two contributions in Eq.~(\ref{eq1aaa})
  cannot interfere if $m_c = 0$.
  For the massive charm quark the interference
  does not vanish and is proportional to the charm mass in the first power. Whether or not the interference contribution
  is negligible depends on the relative magnitude of the two amplitudes in Eq.~(\ref{eq1aaa}). Leading-order computations
 with massless quarks  show that  the charm-Yukawa independent
 amplitude $g_{ggH} {\cal M}_2$ in Eq.~(\ref{eq1aaa}) is larger than the charm-Yukawa dependent one $g_{\yuk} {\cal M}_1$ suggesting
 that the interference  may be non-negligible.

It is straightforward  to calculate  the interference at  leading order in perturbative QCD. Indeed, the
interference requires {\it one} helicity flip on a charm line that connects initial and final states;
this flip is
accomplished by a {\it single} mass insertion. This  implies that one can
compute the interference of the two amplitudes using  massive charm quarks, take the $m_c \to 0$ limit
and account for the first non-vanishing term  proportional to $m_c$. Since we require a  charm jet in the final
state, none of the kinematic invariants of the $cg \to Hc$ process can be small.  Hence, 
once one power of $m_c$ is extracted, the rest of the leading-order
calculation of  the interference contribution can be
performed using the standard approximation of  massless (charm) quarks.  Such calculation, that we describe
in Section~\ref{sec:nlo-qcd}, shows that the leading-order
interference amounts to about ten  percent of the contribution to the $H+{\rm jet}_c$ cross section
that is proportional to the Yukawa coupling squared.

Although the interference contribution is not  large, it is worth thinking about it
at next-to-leading order (NLO) in perturbative QCD since there are reasons to believe that the interference contribution
is {\it perturbatively unstable}, at variance with the two other contributions to $pp \to H+{\rm jet}_c$ cross section.
Indeed,  even if we require an energetic charm jet in the final
state, soft and collinear kinematic configurations lead to logarithmic
sensitivity of the interference to the charm mass $m_c$. Hence, before the $m_c \to 0$  approximation
can be taken, the quasi-singular contributions proportional to
logarithms of $m_c$ have to be extracted from both real and virtual
corrections  to the interference part of  the production cross section.

One may argue that, since the finite charm mass provides  yet another way to regulate collinear
divergences, it is to be  expected  that the  procedure described above will lead to a
familiar picture of (quasi)-collinear factorization of QCD amplitudes. If so,
all $\ln(m_c)$--dependent  terms  should  disappear once infrared safe
cross sections and distributions  are  computed using short-distance quantities, including
conventional parton distribution functions (PDFs).  However, we will show that for the interference contribution
this expectation  is invalid  and  that
 well-known formulas that describe {\it collinear} factorization of mass singularities
are not applicable in that case.   We will also show that the helicity flip
leads to an appearance of {\it soft-quark} singularities that,
interestingly, make jet algorithms logarithmically-sensitive to
$m_c$.

There are two consequences of the above discussion. First, the
problem of estimating the magnitude of the interference contribution to the
production of a Higgs boson in association with a charm jet turns
into an interesting problem in perturbative QCD that borders on such important issues as
soft and collinear QCD factorization
for mass power corrections \cite{Penin:2014msa,Melnikov:2016emg,Liu:2019oav,Liu:2020wbn,Laenen:2020nrt}.
Second, a more complex pattern of this  factorization, as compared to  the canonical
collinear and soft cases \cite{Catani:2000ef},
implies that NLO QCD corrections to leading-order interference are enhanced by up to two powers of a large logarithm
$\ln Q/m_c$  where $Q$ is a typical hard scale  in the process $pp \to H+{\rm jet}_c$.
For this reason NLO QCD corrections to the interference may  be
expected to be significant and it becomes essential to explicitly compute them. This is what we set out to do
in this paper.

The rest of the paper is organized as follows. In the next section we
derive a relation between ${\overline {\rm MS}}$-regulated and mass-regulated
parton distribution functions at ${\cal O}(\alpha_s)$ using the process of
Higgs boson production in $c \bar c$ annihilation.  We use the established relation to
remove ``conventional'' collinear logarithms from NLO QCD corrections
to the interference contribution to the production of Higgs boson in association with  charm jet.
In Section~\ref{sec:quasi-coll} we  discuss
factorization of mass singularities in  the  interference contribution to  $c g \to Hc$ process
and  show that it works differently  as compared
to  the standard  case \cite{Catani:2000ef}.  In Section~\ref{sec:virt} we briefly describe the technical details of the
calculation of NLO QCD corrections to the interference contribution.
In  Section~\ref{sec:nlo-qcd} we  present phenomenological results and discuss the relative
importance of logarithmically-enhanced terms.
We conclude in Section~\ref{sec:end}.
Additional discussion of soft and collinear limits of the interference contributions as well as some relevant soft
integrals can be found in several  appendices.

\section{Matching parton distribution functions}
\label{sec:pdf-scheme}

It is well-known  that  quark masses screen
collinear singularities. For this reason we can think about small quark masses as a particular choice of
a collinear regulator. Since, when describing ``leading-twist'' inclusive partonic
processes, collinear sensitivity either cancels out or is absorbed
into parton distribution functions, it is possible to derive relations
 between parton distribution functions that are used for
computations with nearly massive and strictly massless quarks by requiring that predictions for  physical processes
are independent of a collinear regulator.\footnote{We note that a  derivation of the initial condition
  for the electron structure   function in QED was recently presented  in  Ref.~\cite{Frixione:2019lga}. There
is a strong conceptual overlap of the discussion in that reference and  the computation reported in this section.}

To derive a relation between ``massive'' and ``massless''  PDFs, we start with the production of a
Higgs boson in an annihilation of two massive charm quarks and write the differential cross section as
\begin{align}
{\rm d} \sigma_{pp \to H} = \sum \limits_{ij}^{} \;
\int {\rm d} x_1 \, {\rm d} x_2 \, f^{(m)}_i(x_1) f^{(m)}_j(x_2) {\rm d} \hat\sigma^{(m)}_{ij \to H + X}.
\end{align}
Here $f_i^{(m)}$ are parton distribution functions and the superscript
$m$ implies that all relevant quantities should be computed using
quark masses as collinear regulators. Also,  $\hat\sigma^{(m)}_{ij \to H + X}$ is the partonic differential cross-section.
At leading order $i(j) = c,\bar c$; at higher orders other channels also contribute.

Calculation at leading order in $\alpha_s$ is straightforward since the leading-order
cross section  $ \sigma^{(m)}_{c \bar c \to H}$
has  a regular $m_c \to 0$ limit.  It follows  that at leading order in $\alpha_s$
there is no difference  between
$f_i^{(m)}$ and conventional ${\overline {\rm MS}}$ parton distribution functions, i.e.
$f^{(m)}_i = f^{\overline {\rm MS}}_i$.

The situation becomes more complicated at next-to-leading order where
the charm quark mass screens collinear singularities; hence, our goal is to re-write
the NLO QCD contributions to the cross section $c \bar c \to H +X$ in such a way that logarithms
of $m_c$ are extracted explicitly.

We begin by considering the process
$c(p_1) + \bar c(p_2) \to H + g(p_3)$ and treating  charm quarks as massive.  Kinematic regions that lead to soft
 and (quasi-)collinear singularities are well understood.
The behavior of matrix elements in these limits is described by
conventional factorization formulas~\cite{Catani:2000ef}.  We can
define a hard $m_c$-independent cross section by subtracting the
singular limits.   When the subtracted terms are added back and integrated
over unresolved parts of the $Hg$ phase space, logarithms of the charm mass appear.
This procedure is identical to  methods developed to extract infrared and collinear singularities from real
emission contributions to partonic cross sections. Its application in the present context
allows us to explicitly extract  logarithms of the charm mass.

To organize the calculation, we follow the nested soft-collinear
subtraction
scheme~\cite{Caola:2017dug,Caola:2019nzf,Caola:2019pfz,Asteriadis:2019dte}
which, at next-to-leading order, is equivalent to the FKS
scheme~\cite{Frixione:1995ms,Frixione:1997np}. We use dimensional regularization\footnote{The space-time dimension
  $d$ is parametrized as $d = 4-2\ep$.} to regularize soft singularities and the charm mass
to regularize the collinear ones.   Using notations from Ref.~\cite{Caola:2017dug},
we  write the
partonic cross section for $c(p_1) + \bar c(p_2) \to H + g(p_3)$ as
\begin{align}
\begin{split}
& 2s\cdot{\rm d} \hat\sigma_{c \bar c \to H+g}=    \int [{\rm d}g_3] F_{\rm LM}(1_c,2_{\bar c} ;3_g) \equiv \langle F_{\rm LM}(1_c ,2_{\bar c} ;3_g) \rangle
  = \langle S_3 F_{\rm LM}(1_c,2_{\bar c} ; 3_g) \rangle
  \\
  &  + \langle (C_{31} + C_{32} ) (I - S_3) F_{\rm LM}(1_c,2_{\bar c}; 3_g) \rangle
  + \langle (I- C_{31} - C_{32} ) (I - S_3) F_{\rm LM}(1_{c},2_{\bar c}; 3_g) \rangle.
\end{split}
\label{eq2.2}
\end{align}

The key observation is that since the fully-regulated (last) term in
Eq.~(\ref{eq2.2}) is free of both soft and quasi-collinear
singularities, the limit $m_c \to 0$ can be safely taken there. On the
contrary, both soft and collinear subtraction terms exhibit mass
singularities;  these mass singularities need to be extracted.

Consider the soft limit defined as $\scr{3}{1} \sim \scr{3}{2} \to 0$. It
reads~\cite{Catani:2000ef}
\begin{align}
S_3 F_{\rm LM}(1_c,2_{\bar c}; 3_g) \approx g_s^2 C_F \left (
  \frac{2 (p_1\cdot p_2) }{ (p_1\cdot p_3) (p_2\cdot p_3)}  - \frac{m_c^2}{(p_1 \cdot p_3)^2} - \frac{m_c^2}{(p_2 \cdot p_3)^2}
\right ) F_{\rm LM}(1_c,2_{\bar c}),
\label{eq2.3}
\end{align}
where $g_s$ is the unrenormalized strong coupling constant.

Since the soft gluon  decouples from the function $F_{\rm LM}(1_c,2_{\bar c})$
we can integrate Eq.~(\ref{eq2.3}) over the gluon phase space. 
We work in the center-of-mass frame of the colliding charm  partons  and parametrize their energies  
as  $E_1 = E_2 = E$.  The center of mass energy squared in the
massless approximation is then $s = 4E^2$.  We also cut integrals over
gluon energy at $E_3 = E_{\rm max}$, cf. Ref~\cite{Caola:2017dug}. Integrating Eq.~(\ref{eq2.3}) over gluon phase space
$[{\rm d} g_3]$ and taking the
$m_c \to 0$ limit, we find
\begin{align}
\langle S_3 F_{\rm LM}(1_c,2_c;3_g) \rangle  =
-\frac{C_F \left [ \alpha_s \right ]E_{\rm max}^{-2\ep}}{\ep}
\big[ 2 \; I_{1m}(E) - I_{2m}(E) \big] \langle F_{\rm LM}({\tilde 1_c}, \tilde 2_{\bar c}) \rangle,
\label{eq2.4}
\end{align}
where  $[\alpha_s] = g_s^2 \Omega^{(d-2)}/(2 (2\pi)^{d-1})$ and $\Omega^{(d-2)}$ is the solid angle of the
$(d-2)$-dimensional space.\footnote{The solid angle of the $d$-dimensional space is $\Omega^{(d)} = 2\pi^{d/2}/\Gamma\big(d/2\big)$.}   The notation
$\tilde{1}_c (\tilde{2}_{\bar c})$ implies that the corresponding four-momenta
should be taken in the massless approximation.  The two integrals
$I_{1m(2m)}$ in Eq.~(\ref{eq2.4}) read
\begin{align}
\begin{split}
  &     I_{1m}(E) =
  \int \limits_{-1}^{1} \frac{ {\rm d} \cos \theta ( \sin^2 \theta )^{-\ep}}{1-\beta \cos \theta}
  \approx - \frac{4^{-\ep} \Gamma^2(1-\ep)}{ \ep \; \Gamma(1-2\ep)}
  \left [
    1 - \frac{\Gamma(1+\ep) \Gamma(1-2\ep)}{\Gamma(1-\ep)} \left ( \frac{m_c^2}{4E^2} \right )^{-\ep}  \right ],
  \\
  &     I_{2m}(E) = \frac{m_c^2}{E^2}
  \int \limits_{-1}^{1} \frac{ {\rm d} \cos \theta (\sin^2 \theta )^{-\ep}}{ \left ( 1
      -\beta \cos \theta \right )^2 }
  \approx  2 \left ( \frac{m_c^2}{E^2} \right )^{-\ep} \Gamma(1-\ep) \Gamma(1+\ep),
  \label{eq2.5}
\end{split}
\end{align}
where $\beta = \sqrt{1-m_c^2/E^2}$ and we neglected all power-suppressed
terms when writing the results.

Collinear subtraction terms contain quasi-collinear singularities. The
two collinear limits correspond to two distinct cases,
$\scr{1}{3} \sim m_c^2 \to 0 $ and $\scr{2}{3} \sim m_c^2 \to 0$. They read
\begin{align}
  C_{3i} \; F_{\rm LM}(1_c,2_{\bar c}; 3_g)
  = \frac{g_s^2}{( p_i \cdot p_3 )} \left [ P_{qq}(z)
    - \frac{C_F m_c^2 z}{(p_i\cdot p_3)}  \right ]  \frac{F^{(i)}_{\rm LM}({\tilde 1_c},\tilde 2_{\bar c};z) }{z},
 \label{eq2.6}
\end{align}
where
\begin{align}
F^{(i)}_{\rm LM}({\tilde 1_c},\tilde 2_{\bar c}; z) = \delta_{i1} \; F_{\rm LM}( z \cdot {\tilde 1_c},\tilde 2_{\bar c})
+ \delta_{i2} \; F_{\rm LM}({\tilde 1_c},z \cdot \tilde 2_{\bar c} ),
\end{align}
and
\begin{align}
P_{qq}(z) = C_F \left ( \frac{1 + z^2}{1-z}-\ep(1-z) \right )
\end{align}
is the collinear splitting function. The variable $z$ is defined as
$z = (E_i - E_3)/E_i$ with $i=1,2$,  as appropriate.

Since, when computing the  collinear limits, we do not change the gluon phase
space  \cite{Caola:2017dug}, integrated collinear subtraction terms are still described by
angular integrals $I_{1m(2m)}$ shown  in Eq.~(\ref{eq2.5}). Performing the
soft subtraction of the collinear-subtracted cross section, we find
\begin{align}
\langle C_{31} (I - S_3) F_{\rm LM}(1_c,2_{\bar c}; 3_{g}) \rangle
& = [\alpha_s] E^{-2\ep}
\ \int \limits_{0}^{1} {\rm d} z \; {\cal I}(E,z) \; \left \langle  \frac{F_{\rm LM}(z \cdot {\tilde 1_c}, {\tilde 2_{\bar c}})}{z} \right \rangle ,
\end{align}
where
\begin{equation}
 {\cal I}(E,z) = I_{1m}(E) \;  {\bar P}_{qq}(z)
      - I_{2m}(E)\; {\bar P}_{qq}^{(m)} (z) , 
\end{equation}
and 
\begin{align}
  \begin{split}
     & \bar P_{qq}  = C_F \left (
    \frac{1+z^2}{(1-z)^{1+2\ep}}-\ep(1-z)^{1-2\ep}  + \frac{1}{\ep}\delta(1-z) e^{-2\ep L_1}
  \right ),
  \\
  & \bar P_{qq}^{(m)} = C_F \left (
    \frac{z}{(1-z)^{1+2\ep}} + \frac{1}{2\ep}\delta(1-z) e^{-2\ep L_1}
  \right ),
\end{split}
\end{align}
and $L_1 = \ln(\Emax/E)$.  A similar expression can be written
for $\langle C_{32} (I - S_3) F_{\rm LM}(1,2; 3) \rangle$.

It is straightforward to combine  soft and collinear contributions
and to expand them in $\ep$. At this point, it is convenient to switch
to a strong coupling constant renormalized at the  scale $\mu$. We obtain
\begin{align}
\begin{split}
  \langle  & S_3 F_{\rm LM}(1,2; 3) \rangle
  +\langle C_{31} (I - S_3) F_{\rm LM}(1,2; 3) \rangle
  +\langle C_{32} (I - S_3) F_{\rm LM}(1,2; 3) \rangle
  =
  \\
  & =  C_F \frac{\alpha_s(\mu)}{2\pi}
  \left \{
    -\frac{2L_c}{\ep} - L_c^2 + 1 - \frac{2\pi^2}{3} -3L_c+ 2 L_c \ln \frac{\mh^2}{\mu^2}
  \right \} \langle F_{\rm LM}(1,2) \rangle 
  \\
  &
  + C_F \frac{\alpha_s(\mu)}{2\pi}
  \int \limits_{0}^{1} {\rm d} z \;
  \left \{  \left [ \frac{1+z^2}{1-z} \right ]_+  L_c + (1-z)
  \right \}
  \left \langle    \frac{F_{\rm LM} (z \cdot \tilde 1_c, \tilde 2_{\bar c}) }{z}
    + \frac{F_{\rm LM} ( \tilde 1_c,  z \cdot \tilde 2_{\bar c} ) }{z}
  \right \rangle ,
\end{split}
\end{align}
where $L_c = \ln( \mh^2/m_c^2) - 1$.

To determine  full NLO QCD correction  to  $c \bar c \to H$ cross section,
we need to include virtual corrections.  They are computed in a
standard way (see e.g.  Ref.~\cite{Behring:2019oci} where such a computation is reported);
the result is then expanded around  $m_c = 0$.
We  renormalize the Yukawa coupling in the ${\overline {\rm MS}}$ scheme at
the scale $\mu$.  The result reads
\begin{align}
2s\cdot{\rm d} \hat\sigma_V = C_F \frac{\alpha_s(\mu)}{2 \pi}
\left [
  \frac{2}{\ep} L_c + \frac{4 \pi^2}{3} + 4 + L_c^2 +  ( 2 L_c +3 )\ln \frac{\mu^2}{\mh^2}
   + 3 L_c
\right ] \langle F_{\rm LM}( \tilde 1_c, \tilde 2_{\bar c} ) \rangle .
\end{align}

Upon combining virtual, soft, collinear and fully-regulated terms, we
obtain the following NLO QCD contribution to $c \bar c \to H + X$ cross section
\begin{align}
\begin{split}
  2s\cdot{\rm d} \hat\sigma_{\rm NLO} ={}&  \langle (I- C_{31} - C_{32} ) (I - S_3) F_{\rm LM}({\tilde 1}_{c},{\tilde 2}_{\bar c}; 3_g) \rangle
\\
&   + C_F \frac{\alpha_s(\mu) }{2\pi} \left ( \frac{2 \pi^2}{3} +5 - 3 \ln \frac{\mh^2}{\mu^2} \right )
  \left \langle F_{\rm LM}(\tilde 1_c, \tilde 2_{\bar c}) \right \rangle
  \\
  &
  +  \frac{\alpha_s(\mu) }{2\pi} \sum \limits_{i=1}^{2}  \int \limits_{0}^{1} {\rm d} z \;
  \left \{
    P^{AP}_{qq}(z) \ln \frac{\mu^2}{m_c^2}
    + P_{qq}^{\rm fin}(z)
    \right \} \left \langle \frac{F^{(i)}_{\rm LM}( {\tilde 1}_c , {\tilde 2}_{\bar c}; z )}{z} \right \rangle ,
\end{split}
\label{eq2.13}
\end{align}
where
\begin{align}
  P_{qq}^{\rm AP} ={}& C_F \left [ \frac{1+z^2}{1-z} \right ]_+  &
  & {\rm and} &
  P_{qq}^{\rm fin} ={}&  C_F \left [ \frac{1+z^2}{1-z} \right ]_+ \left ( \ln \frac{\mh^2}{\mu^2} -1 \right )+C_F(1-z)\,.
\end{align}
The first term on the right hand side of Eq.~(\ref{eq2.13}) is the hard inelastic contribution; it can be computed
directly in  the massless limit, $m_c = 0$. The second term on the right hand side of Eq.~(\ref{eq2.13}) is the soft-virtual
piece; it describes kinematic configuration that is  equivalent to the leading-order one. The third term in Eq.~(\ref{eq2.13})
describes kinematic configurations that are boosted relative to the leading-order ones; we note that the residual logarithmic
dependence on $m_c$
is present in these contributions {\it only}.

A similar computation for {\it massless}  charm partons requires, in addition, a
collinear PDF renormalization to make the partonic cross section collinear-finite and independent of the regularization parameter $\ep$.  The
result reads
\begin{align}
\begin{split}
  2s\cdot{\rm d} \hat\sigma^{m_c = 0}_{\rm NLO} ={}&
  \langle (I- C_{31} - C_{32} ) (I - S_3) F_{\rm LM}({\tilde 1}_{c},{\tilde 2}_{\bar c}; 3_g) \rangle
  \\
  & +   C_F \frac{\alpha_s(\mu)}{2 \pi} \left [ 5 + \frac{2 \pi^2}{3} -
    3 \ln \frac{\mh^2}{\mu^2}  \right ] \langle F_{\rm LM}(\tilde 1_c, \tilde 2_{\bar c} ) \rangle 
  \\
  & +\frac{ \alpha_s(\mu) }{2 \pi} \; \sum \limits_{i=1}^{2} \; \int \limits_{0}^{1} {\rm d} z \; P_{qq}^{(\ep)}(z)
  \left \langle \frac{F^{(i)}_{\rm LM}( {\tilde 1}_c , {\tilde 2}_{\bar c}; z )}{z} \right \rangle ,
\end{split}
\label{eq2.15}
\end{align}
where
\begin{align}
P_{qq}^{(\ep)}(z) =  C_F \left [\frac{1+z^2}{1-z} \right ]_+ \ln \frac{\mh^2}{\mu^2} +
2 C_F \left [ \frac{1+z^2}{1-z} \ln(1-z) \right ]_+ + C_F (1-z) .
\end{align}

The partonic cross sections in Eqs.~(\ref{eq2.13}) and (\ref{eq2.15})  should be convoluted with {\it different}
parton distribution functions  to obtain hadronic  cross sections: in case of Eq.~(\ref{eq2.15}) we must
use conventional $\overline {\rm MS}$ PDFs whereas in case when the incoming charm quarks are massive
a special set of PDFs is required.   Nevertheless,
since $m_c$ is just  a collinear regulator, results for
short-distance hadronic  cross sections should be the same, independent of whether one  starts with nearly massive
or massless charm quarks. This
requirement allows us to  derive a relation between the ``massive'' and the $\overline {\rm MS}$ PDFs.  It reads
\begin{align}
  f_{a}^{(m)} = \hat O_{ab} \otimes f^{\MSbar}_b,
  \label{eq2.16}
\end{align}
where
\begin{align}
\hat O_{ab}(z)= \delta_{ab} \delta(1-z)  + \left(\frac{\alpha_s}{2\pi}\right) G_{ab}(z) + \dots.
\end{align}

The computation that we just described allows us to determine the coefficient $ G_{cc}(z)$. We find 
\begin{align}
G_{cc}(z) = -\ln \left ( \frac{\mu^2}{m_c^2} \right ) P_{qq}^{\rm AP}(z)
+ C_F \left [ \frac{1+z^2}{1-z} (1 + 2 \ln(1-z) ) \right ]_+.
\label{eq2.17}
\end{align}

We can also compute ``off-diagonal''coefficients $G_{ab}$ that involve charm quark and gluon  PDFs; they are important for
removing mass singularities  that arise in $g \to c $ and $c \to g$ transitions.  Computations proceed along the same lines as described
above except that we employ  other  partonic processes for the analysis.
Namely, we derive a relation for $g \to c$ transition by considering
a $cg \to Hc$ process in a theory where only Yukawa coupling is present and no direct $ggH$ coupling is allowed. To derive a relation
for $c \to g$ transition, we again consider $cg \to Hc$ process but now only allow for the $ggH$ coupling. In both cases
only (quasi)-collinear singularities are present; this simplifies the required computations significantly.  We find
\begin{align}
\begin{split}
& G_{cg}(z) = -\ln \left ( \frac{\mu^2}{m_c^2} \right ) P^{\rm AP}_{qg}(z),\\
& G_{gc}(z) = - \left [ \ln \left ( \frac{\mu^2}{m_c^2} \right )
                        - 2\ln \left ( z \right )  - 1\right ] P^{\rm AP}_{gq}(z),
  \label{eq2.20}
\end{split}
\end{align}
where
\begin{align}
    P_{qg}^{\rm AP} = T_R \left[ 1 - 2z(1-z) \right],
    \;\;\;\;\; P_{gq}^{\rm AP}  = C_F \frac{1+(1-z)^2}{z}.
\end{align}

The results for the functions $G_{ab}(z)$  reported in
Eqs.~(\ref{eq2.17},\ref{eq2.20}) are important for the calculation of NLO QCD corrections to the interference contribution
to  Higgs boson production in association with a  charm  jet. Indeed, as explained in the Introduction,
to access the interference,  we need to start with the massive incoming charm quarks and carefully study the  massless limit.
Since the charm mass serves as  a collinear regulator, we are  forced   to use parton distribution
functions $f_i^{(m)}$. We then use the relation Eq.~(\ref{eq2.16}) to express these functions through the
conventional $\overline {\rm MS}$
PDFs and, in doing so, remove
logarithms of $m_c$  that are
associated with the  radiation by the incoming charm quarks.
Because  the interference contribution to $pp \to H+{\rm jet}_c$ involves a helicity flip, standard
collinear logarithms associated with initial state emissions  are not the only logarithms of the charm mass that
appear in the cross section.  We elaborate on this statement in the next section.

\section{Interference contributions and factorization in the quasi-collinear limit}
\label{sec:quasi-coll}

The discussion in the previous section shows that the
dependence on $m_c$  disappears from hard cross sections
provided that conventional factorization formulas for soft and
quasi-collinear singularities,  Eqs.~(\ref{eq2.3},\ref{eq2.6}), hold true.  However, since the interference
contribution requires a helicity flip, its soft and quasi-collinear
limits are different from the conventional ones.  As we explain below, such limits can still
be described by simpler matrix elements but these matrix elements do not always correspond to  processes with
reduced multiplicities  of final state particles.

To discuss and illustrate these subtleties in more detail, consider the
process
\begin{align}
  \label{eq:cgHcg-proc}
  c(p_1) + g(p_2)
  \longrightarrow
  H(p_H) + c(p_3) + g(p_4)
  \,.
\end{align}
The first point that needs to be emphasized is that, if we work with a
finite charm mass, {\it true} soft and collinear limits of the process in
Eq.~\eqref{eq:cgHcg-proc} are, in fact, conventional. These
contributions can be extracted and combined with the virtual corrections to $cg \to Hc$
and renormalized gluon parton distribution function giving a  finite
result for the partonic cross section.   Such a procedure is identical to   what is usually done in  NLO
QCD computations \cite{Frixione:1995ms,Frixione:1997np,Catani:1996vz}
that are traditionally performed  using dimensional regularization for  soft and collinear divergences.
However,  cancellation of ``true'' infrared and
collinear divergences does not tell us anything about non-analytic
dependence of  partonic cross sections  on the charm mass   that we need to extract
before taking  the $m_c \to 0$ limit.

In this paper, we  adopt a  pragmatic approach and  extract all the terms that are singular in the
$m_c \to 0$ limit by studying interference contributions to  squares of scattering amplitudes,
computed {\it explicitly } with massive charm quarks, for all  relevant partonic processes,
$cg \to Hcg  \,$,  $gg \to H\cc \,$,
$cq   \to Hcq  \,$,
$cc  \to Hcc  \,$,
$\cc  \to H\cc \,$.
In that sense,  we do not attempt  to develop an  understanding of infrared and quasi-collinear factorization
for  generic processes  that involve a helicity  flip.  However, to illustrate main  differences with the conventional
collinear factorization we discuss an   emission of a collinear gluon
off an incoming charm  quark
in case of the interference contribution in some detail.

To this end, we consider the  process in Eq.~(\ref{eq:cgHcg-proc})
in the quasi-collinear limit   $p_1 \cdot p_4 \sim m_c^2$.  To describe this limit, we divide  the
amplitude for the full process $cg \to Hcg$ into two parts
\begin{align}
{\cal M} = {\cal M}_{\rm sing} + {\cal M}_{\rm fin},
\label{eq3.1}
\end{align}
where ${\cal M}_{\rm sing}$ refers to diagrams where the gluon is emitted
off the incoming quark with the momentum $p_1$ and ${\cal M}_{\rm fin}$ refers to the remaining diagrams.
The first contribution
(${\cal M}_{\rm sing}$) is singular in the
$\scr{1}{4} \sim m_c^2 \to 0$ limit whereas the second one
(${\cal M}_{\rm fin}$)
is not.

Upon squaring the amplitude, Eq.~(\ref{eq3.1}), and summing over
polarizations of initial and final state particles, we obtain
\begin{align}
\sum  \limits_{\rm pol}^{} |{\cal M}|^2 = \sum \limits_{\rm pol}^{} |{\cal M}_{\rm sing}|^2
+ \sum \limits_{\rm pol}^{} \left ( {\cal M}_{\rm sing} {\cal M}_{\rm fin}^{\dagger}+{\rm h.c.} \right )
+\ldots\,,
\label{eq3.2}
\end{align}
where the ellipsis stands for contributions that are finite in the
$\scr{1}{4} \sim m_c^2 \to 0$ limit.  We note that the product of
singular and non-singular contributions to the amplitude squared that we retain in
Eq.~(\ref{eq3.2}) is known to be non-singular  in the conventional
quasi-collinear limits provided that physical polarizations are used
to describe the emitted gluon.  As we will show  below, this is {\it not the case}
for the interference contributions considered in this paper.

To  extract the quasi-collinear singularities from the
square of the amplitude in  Eq.~(\ref{eq3.2}) we need to analyze the quasi-collinear kinematics;
for this analysis there is no difference between helicity-conserving and
helicity-flipping  contributions.  Indeed, following the standard approach,  we re-write the
four-momentum of the incoming charm quark and the four-momentum of the
emitted gluon through massless momenta ${\tilde p}_1$ and $p_2$ and find
\begin{align}
p_1 = \left ( 1 - \frac{m_c^2}{s} \right ) {\tilde p}_1 + \left(\frac{m_c^2}{s}\right)  p_2 +{\cal O}(m_c^4)\,,\;\;\;\;\;\;\;
p_4 = (1-z) \tilde p_1 + y p_2  + p_{4,\perp}.
\label{eq3.4aaaa}
  \end{align}
In the above equation,  $s = 2 \tilde p_1 \cdot p_2$ and
$p_{4,\perp} \cdot \tilde p_1 = p_{4,\perp} \cdot p_2 = 0$.  We use
the on-shell condition $p_4^2 = 0$ and obtain
\begin{align}
y = -\frac{p_{4,\perp}^2}{(1-z)s}.
\end{align}
It follows that
\begin{align}
2 p_1\cdot p_4 \approx s \left ( (1-z) \frac{m_c^2}{s} +  y \right )
=  \frac{1}{1-z} \left ( (1-z)^2 m_c^2 - {p}_{4,\perp}^2 \right ).
\end{align}
We conclude that  the kinematic region where  $p_{4,\perp}^2 \sim m_c^2$  provides
unsuppressed contributions to the cross section in the $m_c \to 0$ limit.

We write the singular contribution as follows
\begin{align}
{\cal M}_{\rm sing} = -g_s t^{a_4}_{ i_c i_1}\hat {\cal A}^{i_c}_{\rm sing}
\; \frac{\hat p_1 - \hat p_4 + m_c}{2( p_1 \cdot p_4)}  \; \hat \epsilon_4 \; u(p_1),
\end{align}
and use the decomposition of the four-momenta $p_{1,4}$ given in Eq.~(\ref{eq3.4aaaa})  to obtain
\begin{align}
{\cal M}_{\rm sing} =
\frac{g_s t^{a_4}_{ i_c i_1}}{2 (p_1\cdot p_4)} \hat {\cal A}^{i_c}_{\rm sing} \left [ -\frac{2 p_{4,\perp} \cdot \epsilon_4}{1-z}
  + \hat \epsilon_4 \left ( m_c (1-z) + \hat p_{4,\perp} + \kappa \hat p_2 \right )
\right ] u(p_1)\,.
\label{eq3.7}
\end{align}
In Eq.~(\ref{eq3.7}), we introduced a parameter $\kappa$ defined as
\begin{align}
  \kappa = -\frac{m_c^2(1-z)}{s} - \frac{p_{4,\perp}^2}{s (1-z)}.
  \label{eq3.8}
\end{align}
We note that in deriving Eq.~(\ref{eq3.7}) we have used
$p_{4} \cdot \epsilon_4 = 0$ and the gauge fixing condition
$p_2 \cdot \epsilon_4 = 0$.

The result for the singular contribution, Eq.~(\ref{eq3.7}), is generic;  it does not distinguish between helicity-conserving and helicity-flipping
contributions. However, it is easy to see that there is a difference between the two.
For example, since the helicity-flipping contribution requires one
{\it additional} power
of $m_c$, one can  convince oneself  that a combination of terms labeled
as $\kappa$ in Eq.~(\ref{eq3.8}) may contribute to the collinear limit of the interference but it cannot
contribute to the collinear limit of the helicity-conserving amplitudes.

Hence, performing standard manipulations and paying attention to subtleties indicated above, we obtain
the contribution to the interference that is non-analytic  in the $m_c \to 0$ limit. It reads
\begin{align}
\begin{split}
  &  \lim_{p_1\cdot p_4 \to 0}  {\rm Int} \left [ {\cal M}^2(1_c,2_g;3_c,4_g)    \right ]
   = g_s^2  \Bigg [ \left ( \frac{P_{qq}(z) }{(p_1\cdot p_4)}
    - \frac{ C_F m_c^2 \, z }{(p_1\cdot p_4)^2} \right )
  {\rm Int} \left [  \frac{|{\cal M}(z 1_c,2_g;3_c)|^2}{z} \right ]
  \\
  &
  -  \frac{C_F m_c (1-z)}{z  (p_1\cdot p_4) } {\rm Int} \left [  {\rm Tr} \left [ \hat {\cal A}^{i_c}(z \cdot \tilde 1_c,2_g; \tilde 3_c)
      \hat {\cal A}^{i_c,\dagger}(z \cdot \tilde 1_c,2_g; \tilde 3_c) \right] \right ]
  \Bigg ]
  \\
  &
  + g_s C_F  \; \frac{(1-z) m_c }{2 (p_1\cdot p_4)} {\rm Int} \left [
    {\rm Tr} \left [   \hat p_1  {\cal A}^{i_c}(z \cdot \tilde 1_c,2_g;\tilde 3_c)
      \hat {\cal A}^{i_c,\dagger}_{\rm fin}(\tilde 1_c,2_g;\tilde 3_c,(1-z) \cdot \tilde 1_g ) \hat \epsilon_4 + {\rm h.c.}
    \right ] \right ].
\end{split}
\label{eq3.10}
\end{align}
It is understood that ${\rm Int} [...]$ extracts  the interference
contributions from the relevant quantities;  sums over colors and
polarizations are  implicit. The quantities  ${\cal A}$ are related to amplitudes ${\cal M}$ in the following way
\begin{align}
  {\cal M}(\tilde 1_c,2_g; \tilde 3_c) = \hat {\cal A}^{i_1}(\tilde 1_c,2_g; \tilde 3_c) \; u(\tilde p_1),
  \;\;\;\;
      {\cal M}_{\rm fin}(\tilde 1_c,2_g;\tilde 3_c,4_g  ) = \hat {\cal A}^{i_1}_{\rm fin}(\tilde 1_c,2_g;\tilde 3_c,4_g ) \; u(\tilde p_1).
      \label{eq3.12}
\end{align}
They have to be computed in the $m_c = 0$ limit.\footnote{We remind the reader that  the notation
  $\tilde i$ implies that a light-cone four-momentum
  of a particle $i$ must be used in the computation.}

It is instructive to discuss the origin of the different terms in Eq.~(\ref{eq3.10}).
The first term on the right-hand side of  Eq.~(\ref{eq3.10}) contains
the leading-order interference multiplied with the standard massive
collinear splitting function; this is the  conventional quasi-collinear limit applied to the interference.
If only this term were present in Eq.~(\ref{eq3.10}), there would be no differences in the collinear factorization
between  helicity-changing and helicity-conserving contributions.

The second and the third terms on the right-hand side of Eq.~(\ref{eq3.10}) are  new structures; they
appear because  the required  helicity flip can occur on the {\it external charm quark lines}.  To illustrate this point,
we square  the first equation in Eq.~(\ref{eq3.12}), sum over polarizations and find
\begin{align}
\sum \limits_{\rm spins} |{\cal M}(1_c,2_g;3_c)|^2 = {\rm Tr} \left [ ( \hat p_1 + m_c) \hat {\cal A}^{i_c}
\hat {\cal A}^{i_c,\dagger} \right ].
\end{align}
The interference requires a helicity flip that is facilitated by a single mass insertion.
The above equation shows that this mass insertion can occur
either in the $(\hat p_1 + m_c)$  density matrix of the external quark or ``inside'' the
$\hat {\cal A} \hat {\cal A}^\dagger$ term. The structure that appears in
the second term on the right hand side of  Eq.~(\ref{eq3.10}) originates from the mass term in
the density matrix. Once the mass term is extracted, the rest can be computed in the massless approximation. We find
\begin{align}
  \begin{split}
& {\rm Int} \left [ {\rm Tr} \left [ ( \hat p_3 + m_c) \hat {\cal A}^{i_c}(1_c,2_g;3_c)
      \hat {\cal A}^{i_c,\dagger}(1_c,2_g;3_c) \right ] \right ]\to
    \\
& m_c \; {\rm Int} \left[ {\rm Tr} \left [ \hat {\cal A}^{i_c}(\tilde 1_c,2_g;\tilde 3_c)
        \hat {\cal A}^{i_c,\dagger}(\tilde 1_c,2_g;\tilde 3_c) \right ]
      \right ].
\end{split}
\end{align}

The last term on the right hand side in Eq.~(\ref{eq3.10}) describes a
quasi-collinear singularity that originates from the interference of singular and
regular contributions in Eq.~(\ref{eq3.2}); it is  particular to the
helicity-flipping  case  and does not appear in  the conventional collinear limits. As a consequence, this contribution
still depends on the part of the reduced matrix element of
the original $2 \to 3$ process calculated in the  strict collinear limit of the incoming massless charm quark and the emitted gluon.

We emphasize once again  that Eq.~(\ref{eq3.10})  shows clear differences
 between conventional factorization of quasi-collinear
singularities and the factorization in  case of the  helicity flip. These
differences lead to a peculiar structure of the logarithms of the charm quark
mass in the interference contribution as  they  do not follow canonical pattern and cannot be
removed by a transition to  ${\overline {\rm MS}}$ parton distribution functions.
In addition, we also find that the interference contributions exhibit {\it quasi-soft} quark singularities that also lead
to logarithms of the charm mass.   Although it would be interesting to understand  factorization of mass singularities in
 processes with the helicity flip  from a more general perspective,
our strategy for now  is to {\it explicitly} compute all  relevant contributions within fixed-order perturbation theory extracting
all non-analytic $m_c$-dependent terms along the way. Additional details of our approach
can be found in several  appendices.

\section{Technical details of the calculation}
\label{sec:virt}
In this section we briefly describe  calculation of  one-loop
 and real emission contributions to the interference part of the $cg \to Hc$ process.
We begin with the discussion of the virtual corrections.

We compute the relevant one-loop diagrams keeping  charm-quark masses finite.  We employ  the standard Passarino-Veltman
reduction~\cite{Passarino:1978jh} to express  the $cg \to Hc$ amplitude in terms of one-loop scalar integrals.\footnote{We use \texttt{FeynCalc}~\cite{Mertig:1990an,Shtabovenko:2020gxv} for a cross-check of our computation.}
After computing the one-loop contribution to the interference, we expand the expression around  $\mc =  0$
and keep the leading ${\cal O}(m_c)$ term in this  expansion.\footnote{We
  employed  the Package-X program \cite{Patel:2015tea} for numerical checks of scalar integrals and their $\mc\to0$ expansion.}
  The one-loop amplitudes contain  ultraviolet and infrared singularities. The
ultraviolet singularities are removed by the renormalization. We closely
follow  the discussion in   Appendix~A of   Ref.~\cite{Behring:2019oci}
where  many of the required renormalization constants are presented. Similar to Ref.~\cite{Behring:2019oci},
we renormalize the charm-quark mass in the on-shell scheme but employ the  ${\overline {\rm MS}}$ renormalization for the Yukawa coupling constant.
In addition to the discussion in that reference,
we require the one-loop renormalization constant of the
effective $\ggh$ vertex that we take from  Ref.~\cite{Mondini:2020uyy}.
After the ultraviolet renormalization is performed, the $cg \to Hc$ amplitude still contains $1/\ep$ poles of infrared origin.
These poles satisfy the Catani's formula \cite{Catani:1998bh} and  cancel with similar poles in
real emission contributions to  the partonic cross section.

According to the discussion in Section~\ref{sec:quasi-coll},
factorization of quasi-collinear and quasi-soft singularities in the interference
contribution is not canonical.
This implies that even if  we take the $\mc \to 0$ limit and switch
to $\MSbar$ parton distribution functions, the NLO QCD corrections to the
interference  still  contain logarithms of the charm mass.
Since it is currently unknown how these logarithms can be resummed, we
follow a pragmatic approach.
Namely, we compute relevant virtual and real emission contributions,
extract from them logarithms of the charm mass and take the
$m_c \to 0$ limit once the mass logarithms have been extracted.  To
accomplish this, we construct subtraction terms for the real emission
contributions for soft, collinear, quasi-collinear and quasi-soft singularities by
direct inspection of the relevant matrix elements.  The subtraction terms are then
integrated over unresolved real emission phase space and combined with the PDF
renormalization, including the transition from ``massive'' to $\overline {\rm MS}$ PDFs,
and the virtual corrections.
The only difference with respect to the canonical procedure for NLO
QCD computations is that in our case the subtraction terms are
directly obtained from the squared amplitude and are not written in
terms of easily recognizable  universal functions;
see  Section~\ref{sec:quasi-coll} and Appendix~\ref{sec:subt} for further details. In
particular, even contributions that are enhanced by {\it two} powers of a
logarithm of the charm mass, ${\cal O}(\alpha_s \ln^2(m_c))$, do not appear to be
proportional to the leading-order interference contribution to the cross
section.

\section{Numerical results}
\label{sec:nlo-qcd}

To present numerical results we consider proton-proton collisions at
$13~\tev$.  We take $\mh = 125~\gev$ for the Higgs-boson mass and
$\mc = 1.3~\gev$ for the pole mass of the charm quark.  The charm
Yukawa coupling is calculated using the $\MSbar$ charm mass, 
$\overline{m}_c(\mh) = 0.81~\gev$.\footnote{We use program
  \texttt{RunDec}~\cite{Chetyrkin:2000yt,Herren:2017osy} to compute
  the value of the running charm quark mass.}  We use
\texttt{NNPDF31\_lo\_as\_0118} and \texttt{NNPDF31\_nlo\_as\_0118}
parton distribution functions~\cite{Buckley:2014ana,Ball:2017nwa} for leading and next-to-leading order
computations, respectively.  The value of the strong coupling constant
$\alpha_s$ is calculated  using dedicated routines provided with NNPDF
sets.

To define jets we use standard anti-$k_\perp$ algorithm with
$\Delta R = 0.4$; charm jets are required to contain at least one $c$
or $\bar c$ quark.
For numerical computations we require at least one charm jet with
$\ptj > 20~\gev$ and $|\etaj| < 2.5$. Moreover, we demand that the
charm parton inside the charm jet carries at least $75\%$ of the jet's
transverse momentum.\footnote{If  more than one $c$ or $\bar c$ parton
  is clustered into a jet, we apply this requirement to the
  hardest of them.}
The latter requirement removes kinematic cases where a soft charm is
clustered together with a hard gluon into a charm jet in spite of a large angular separation between the two.
Since, as we explained earlier, our calculation is logarithmically
sensitive to {\it soft emissions of charm quarks}, defining charm jets
with an additional cut on the charm quark transverse momentum allows
us to avoid jet-algorithm dependent  logarithms of $m_c$
that  may  appear otherwise.
We note that we apply  all these requirements even in the subtraction terms where $c$ and $\bar c$
momenta are  computed in the  collinear and/or soft approximations.

\begin{figure}
  \centering
  \begin{subfigure}{0.49\textwidth}
    \centering
    \includegraphics[width=0.95\textwidth]{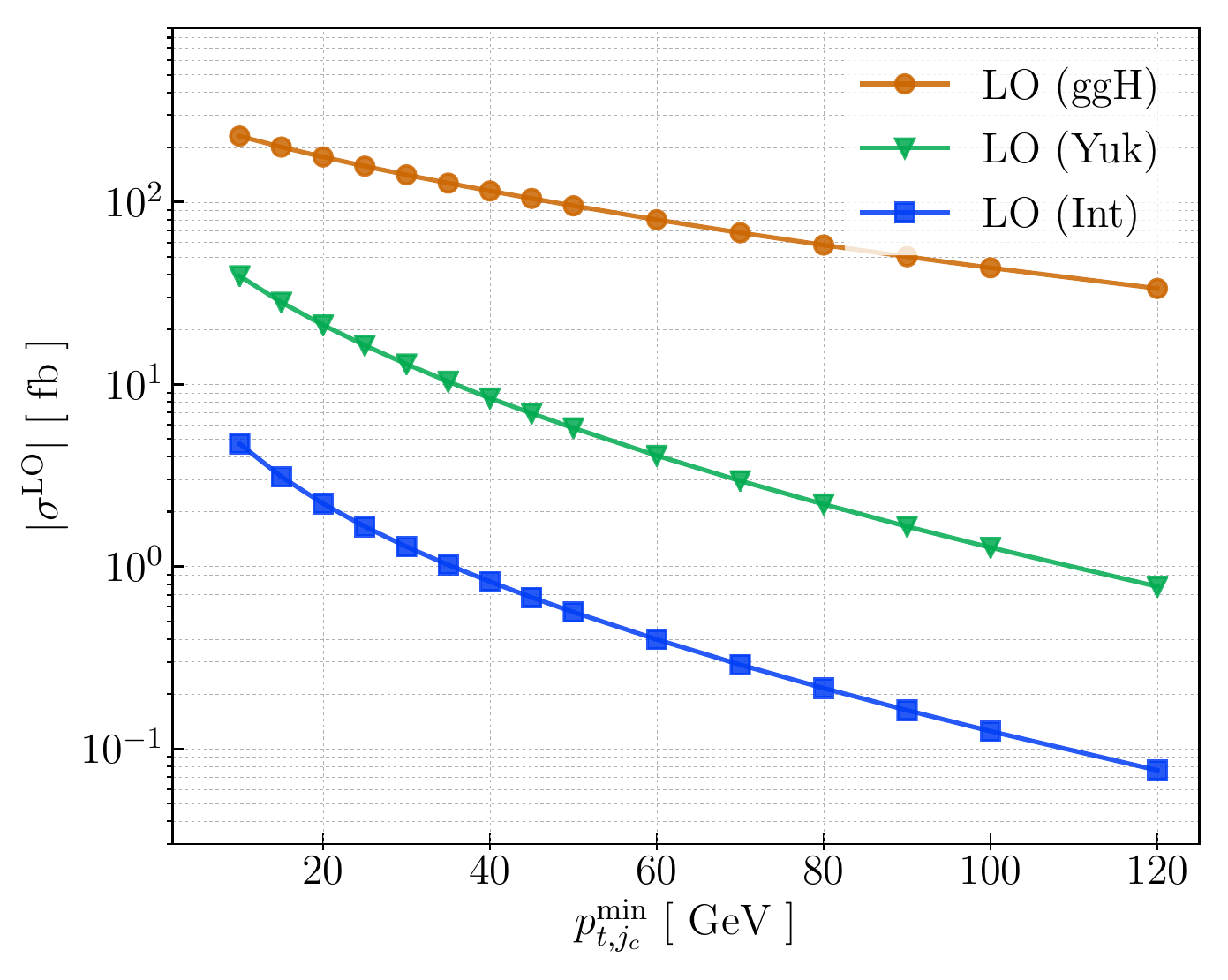}
    \caption{Fiducial cross section (LO)}
    \label{fig:xsec-lo-ptc}
  \end{subfigure}
  \begin{subfigure}{0.49\textwidth}
    \centering
    \includegraphics[width=0.95\textwidth]{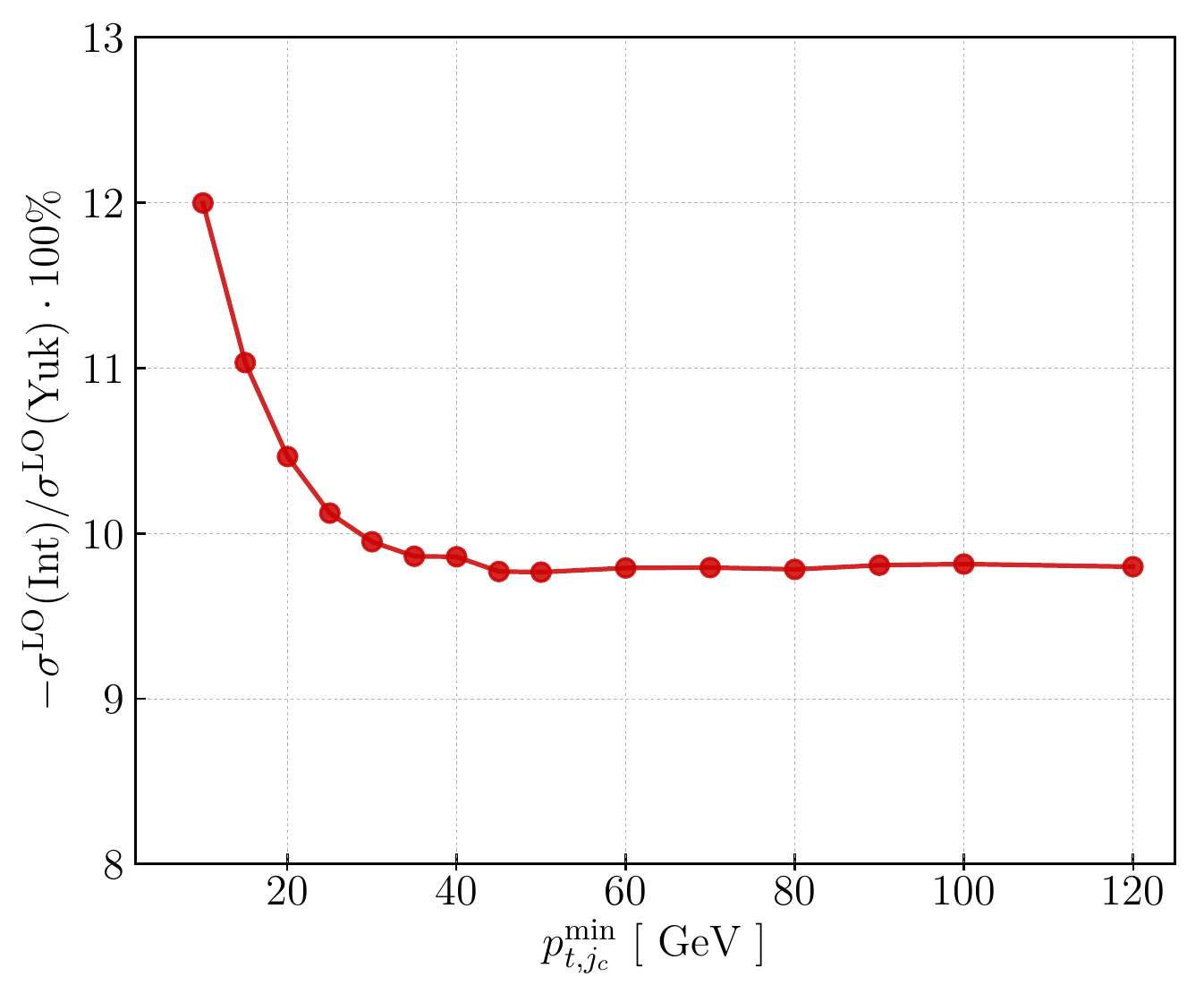}
    \caption{Interference/Yukawa ratio}
    \label{fig:xsec-lo-ptc-ratio}
  \end{subfigure}
  \caption{Leading-order cross sections computed for different values of the charm-jet $p_t$-cut.
    We show  Yukawa-like (green) and
    $ggH$-like (orange) contributions as well as the  {\it absolute value} of the interference (blue). In the right
    pane the ratio of the interference and
    the Yukawa-like fiducial cross section are shown.}
  \label{fig:xsec-lo}
\end{figure}

We start by presenting fiducial cross sections for the three terms in
Eq.~(\ref{eq2aaa}) separately.  Central values for all the cross
sections presented below correspond to the renormalization and
factorization scales set to $\mu_F = \mu_R = \mu = \mh$; subscripts and superscripts
indicate shifts in central values if $\mu = \mh/2$ and $\mu=2 \mh$ are
used in the calculation.  At leading order, we find
\begin{align}
  \sigma^{\rm LO}_{\ggh }    ={}&  176.6^{+47.6}_{-36.5}~\fb\,, &
  \sigma^{\rm LO}_{ \yuk }    ={}&  21.22^{+1.47}_{-1.67}~\fb\,, &
  \sigma^{\rm LO}_{ \interf } ={}&  -2.21^{+0.29}_{-0.31}~\fb\,, &
  \label{eq5.1}
\end{align}
for the $ggH$-dependent cross section, the Yukawa-dependent cross
section and the interference, respectively.
In Figure~\ref{fig:xsec-lo} we show the comparison between $\mu = \mh$
cross sections and the interference in dependence on the cut of the charm jet transverse momentum.
We observe that  the ratio of the interference to the Yukawa-dependent  contribution is about ten percent for
all values of the $p_{t,j}$-cut.

At next-to-leading order the fiducial cross section for the interference term becomes
\begin{align}
  \sigma^{\rm NLO}_{\interf} ={}& -1.024(5)^{+0.224}_{-0.144}~\fb\,.
  \label{eq5.2}
\end{align}
It follows from Eqs.~(\ref{eq5.1},\ref{eq5.2}) that the NLO QCD
corrections decrease the absolute value of the leading-order interference by about fifty 
percent. The scale uncertainty appears to be reduced by about a factor 2.
The NLO result for the interference is
outside the leading-order scale-uncertainty interval, c.f.  Eqs.~(\ref{eq5.1},\ref{eq5.2}),  emphasizing the fact that 
the appearance of the logarithms of the charm
mass in NLO QCD corrections to the interference
makes  the scale variation uncertainty of the leading-order result a very poor  indicator  of
the theoretical uncertainty in this case.

\begin{table}
  \begin{center}
    \begin{tabular}{ c | r r r r r r | r  }
      $\Delta \sigma^{\rm NLO}~[~\fb~]$ & $cg$ & $cq$ & $gg$ & $cc$ & $\cc$ & PDF & sum \\
      \hline
      \emph{const} & $ -1.63 $ & $   0.13 $ & $   2.33 $ & $  0.01 $ & $ -0.01 $ & $  0.11 $ & $  0.94 $\\
      $L$          & $  2.23 $ & $    -   $ & $  -6.33 $ & $ -0.04 $ & $  0.01 $ & $  1.66 $ & $ -2.47 $\\
      $L^2$        & $ -0.06 $ & $    -   $ & $   2.66 $ & $  0.01 $ & $ -0.08 $ & $    -  $ & $  2.52 $\\
      \hline
      total        & $  0.54 $ & $   0.13 $ & $  -1.34 $ & $ -0.02 $ & $ -0.08 $ & $  1.76 $ & $  1.00 $\\
    \end{tabular}
  \end{center}
  \caption{The NLO QCD corrections to the interference split according to partonic channels. The
    results  are given in femtobarns. The column marked ``PDF''
    refers to the PDF-scheme change discussed in
    Section~\ref{sec:pdf-scheme}.  For each partonic channel we show
    ${\cal O}(L^2)$, ${\cal O}(L)$ and ${\cal O}(L^0)$ contributions
    where $L = \ln(\mh/\mc)$.}
    \label{tab:xsec-separation}
\end{table}

It is instructive to  separate the NLO contributions to the interference into 
parts that are independent of $\mc$ and parts that are
logarithmically enhanced  for all the partonic channels.
The relevant results are shown in
Table~\ref{tab:xsec-separation}.
We find that the largest contribution at NLO comes from the
gluon-gluon channel which is enhanced by the large gluon luminosity. Also,
 the charm-gluon ($cg$) and
 the charm-quark channels ($cq$)  provide relatively large contributions.\footnote{Here, by ``quark'' we
   mean any quark of a flavor other than $c$. There is a subtlety related to the $b$-quark
   contribution because  $b$-quarks  have  stronger interactions with Higgs bosons as  compared to charm quarks. 
   Such contributions can, presumably,  be dealt with using $b$ anti-tagging.
When presenting results for the interference we decided to  include contributions of bottom
   quarks, setting bottom Yukawa coupling to zero, 
   but we did check that the flavor-excitation topologies with $b$ in the
   initial state change the results for $(cq)$ channel by about three percent only. }    Note that the $(cq)$ channel is
free of logarithmic contributions since there are no singular limits
that  involve  charm quarks. Contributions related to the PDF
transformation  do not feature the double-logarithmic part since the
${\cal O}(\ln^2 m_c)$ terms originate exclusively from soft-collinear limits that
involve $c$-quarks.

It follows from
Table~\ref{tab:xsec-separation} that double-logarithmic  terms and single-logarithmic terms 
provide  nearly equal, but opposite in sign, contributions to the NLO QCD interference. This cancellation
between terms with different parametric dependence on $m_c$ should be considered as an artifact but it does
emphasize that studying {\it only} the leading logarithmic ${\cal O}(\ln^2 m_c)$ contribution  in this
case is  insufficient for phenomenology. 
We also note that the ${\cal O}(\ln^2 m_c)$ term in the $cg$ channel is quite small reflecting
the fact that there is a very strong -- but incomplete --
cancellation between double-logarithmic  contributions to real and virtual corrections in this case.
Finally, we emphasize that it is unclear to what extent
these various cancellations persist in higher orders;  for this reason, a resummation of charm-mass
logarithms for the interference contribution is desirable.

\begin{figure}
  \centering
  \begin{subfigure}{0.49\textwidth}
    \centering
    \includegraphics[width=0.95\textwidth]{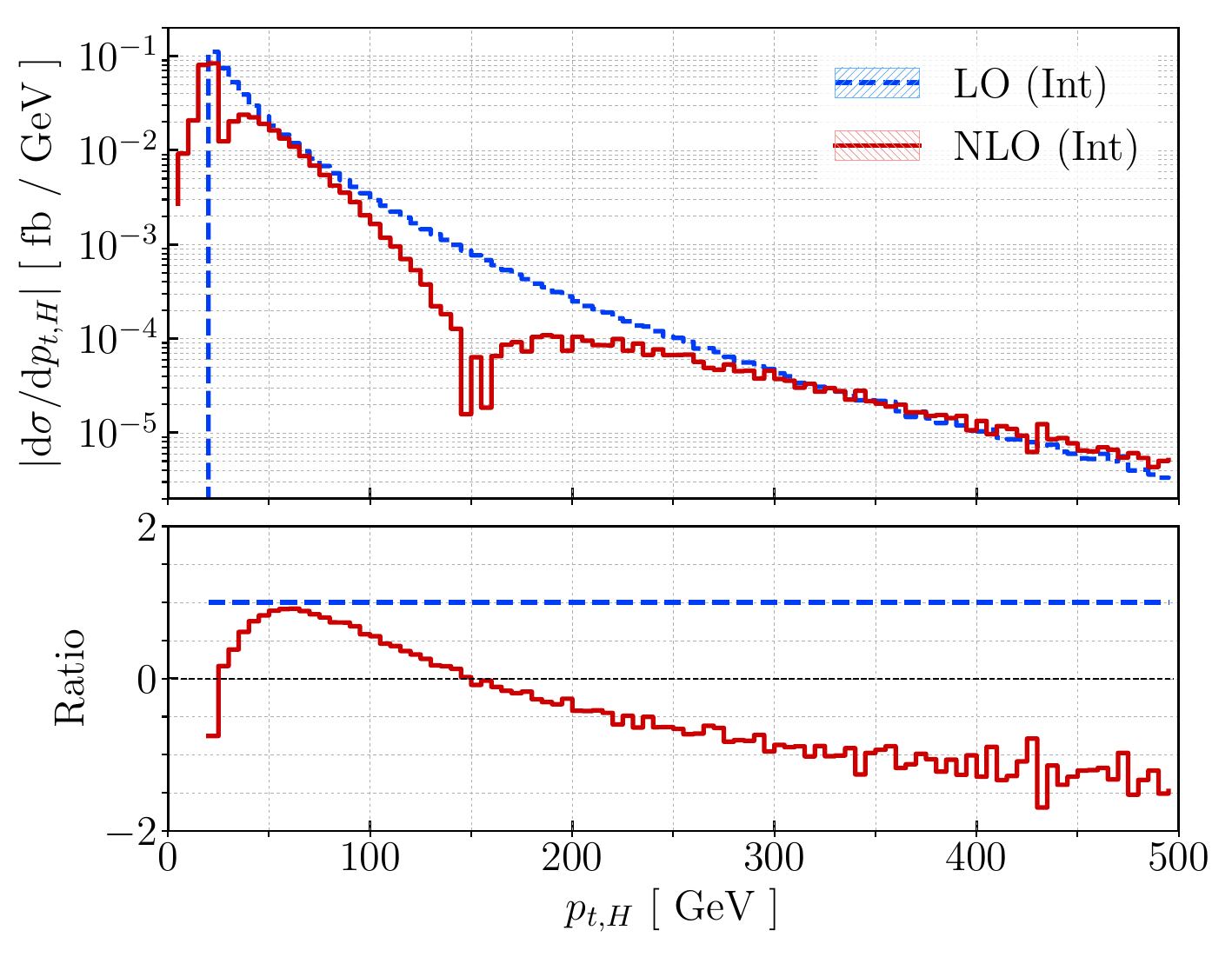}
    \caption{Higgs-boson transverse momentum}
    \label{fig:plot-pth}
  \end{subfigure}
  \begin{subfigure}{0.49\textwidth}
    \centering
    \includegraphics[width=0.95\textwidth]{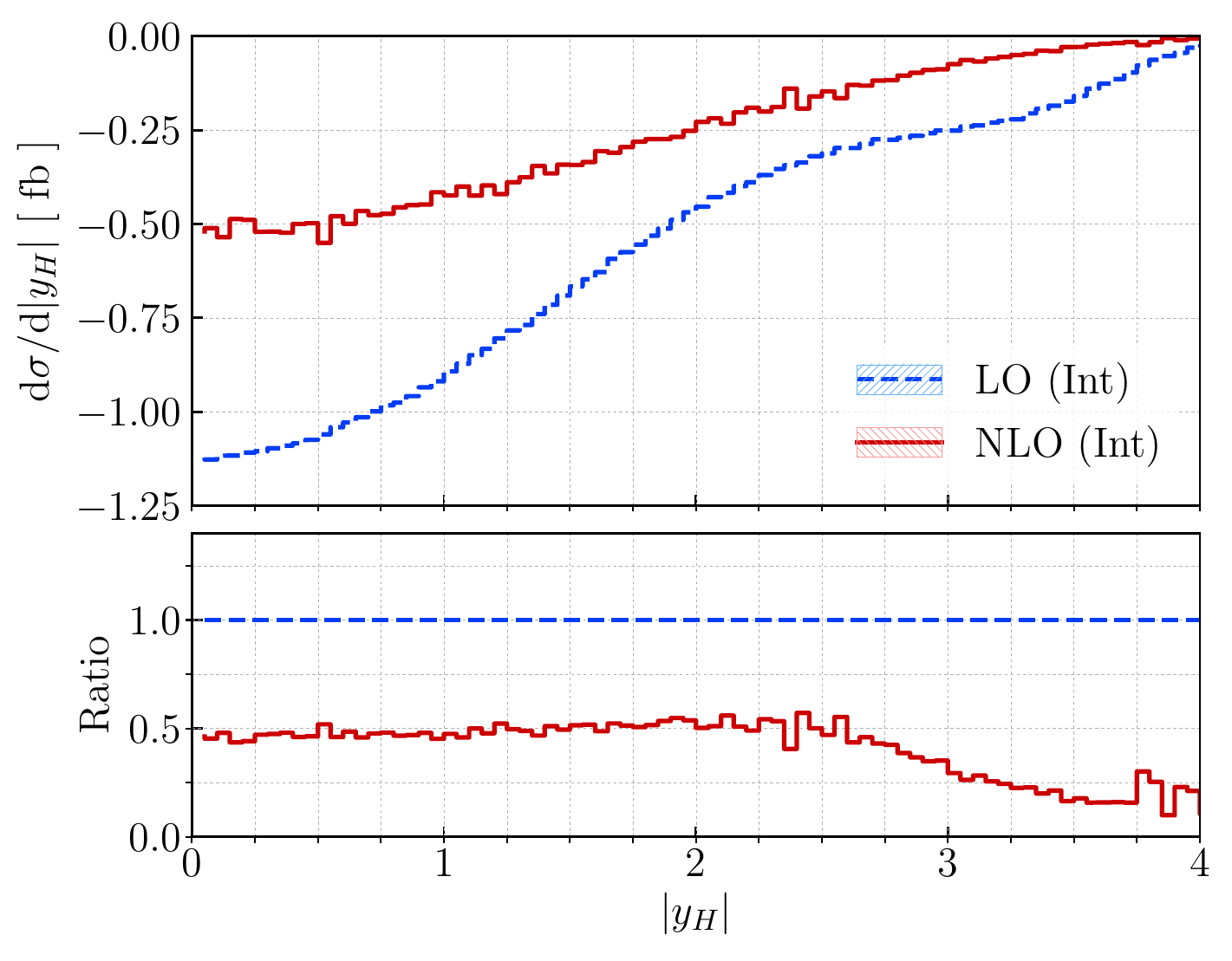}
    \caption{Higgs-boson rapidity}
    \label{fig:plot-yh}
  \end{subfigure}
  \caption{The transverse momentum and rapidity distributions of the Higgs boson
     calculated at LO (blue) and NLO (red) for central
     scale choice. We only consider the interference contribution.  We note that
     the {\it absolute value} of ${\rm d} \sigma_{\rm Int}/{\rm d}p_{t,H}$ is displayed in the left panel. This implies that
     this distribution actually changes sign at around  $p_{t,H} \sim 150~{\rm GeV}$.
     The    lower panels show ratios to the LO interference contribution.}
  \label{fig:plots-nlo}
\end{figure}

We continue with the discussion of kinematic distributions. 
We focus  on the transverse momentum and the rapidity distributions of
Higgs bosons in the interference contribution to $pp \to Hc$ cross
section. They are shown in Figure~\ref{fig:plots-nlo}.
We first discuss the transverse momentum distribution,
Figure~\ref{fig:plot-pth}; when interpreting this figure it is
important to recall that the {\it absolute} value of both LO and NLO distributions
is plotted there and that the LO distribution is always {\it negative}.
We observe in Figure~\ref{fig:plot-pth} that the  leading-order distribution
(blue) is large and negative at small $p_{t,H}$; as $p_{t,H}$ increases, the distribution goes to zero.
The  NLO QCD corrections affect the shape of the $p_{t,H}$ distribution.   Indeed, a  sharp edge at $p_{t,H}=20$~{\rm GeV}, present
at leading order,  gets smeared at NLO.
At  moderate values of transverse momenta $p_{t,H} \sim 60~{\rm GeV}$
the $K$-factor is equal to one, while there is a large ${\cal O}(+50\%)$    reduction\footnote{``Reduction'' in this case means that
  the distribution becomes less negative.} at  
$\pth \sim 100~\gev$. Second, at $p_{t,H} \sim 150~{\rm GeV}$ the NLO distribution goes through zero
and becomes positive for larger values of $p_{t,H}$. Asymptotically, at even higher
$p_{t,H}$ the LO and NLO distributions appear
to be equal in absolute values but {\it opposite} in sign. Of course, all this happens  at such high
values of $p_{t,H}$ that are  irrelevant for phenomenology,  but it is quite a peculiar  feature nevertheless.

Compared to Higgs transverse momentum  distribution, the rapidity distribution of the Higgs boson in the interference contribution 
is much less volatile. Indeed,  it follows from Figure~\ref{fig:plot-yh} that the difference between leading
and next-to-leading-order distributions  is well-described by a constant $K$-factor all the way up to  $|y_H| \sim 2$.
Beyond this value of the rapidity, the NLO distribution goes to zero faster than the LO one. 

\section{Conclusions}
\label{sec:end}

Production of  Higgs bosons in association
with charm jets at the LHC is  mediated by two distinct  mechanisms, one that involves the
charm Yukawa coupling and the other one that involves  an effective $\ggh$
vertex. Their interference  involves a helicity flip and, for this reason,
vanishes in the limit of massless charm quarks.

Since partonic cross sections are routinely computed for massless incoming partons and since the charm quark appears
in the initial state in the main process $cg \to Hc$, it is interesting  to understand how to circumvent the problem
of having to deal with a massive parton in the initial state and to provide reliable  estimate of the interference contribution.

We  have addressed this problem by studying  the $m_c \to 0$ limit of the helicity-flipping interference contribution including NLO QCD corrections.
We have shown that the factorization of quasi-collinear and quasi-soft singularities in this case differs from the canonical
pattern.  We used  explicit expressions for real and virtual matrix elements
to extract logarithms of the charm quark mass and,  having accomplished this, 
took the $m_c \to 0$ limit in the remaining parts of the computation.
We removed parts of the  ${\cal O}(\ln m_c)$ contributions by expressing results 
through  conventional ${\overline {\rm MS}}$ parton distribution functions valid for massless partons.
Nevertheless, given an
unconventional behavior of the interference in quasi-soft and quasi-collinear limits,
logarithms of the charm quark mass survive in the final result for the NLO QCD corrections.

We have found that the absolute value of the leading-order interference is reduced  by about  fifty  percent
once  NLO QCD corrections are accounted for.  This significant but still ``perturbatively acceptable''
reduction   is the result of a
{\it very strong} cancellation between
terms that involve double  and single logarithms of the charm quark mass. 
We have  observed that the NLO QCD corrections
to the interference are  kinematics-dependent and may change shapes of certain kinematic distributions in a significant way.

Higgs boson production in association with a charm jet is a promising way to
study charm Yukawa coupling at the LHC \cite{Brivio:2015fxa}.  The interference contribution, that is estimated
to be about $10$ percent of the Yukawa contribution at leading order, could have been perturbatively unstable given
the required helicity flip and an unconventional  pattern of quasi-soft and quasi-collinear limits.  We addressed
this question by performing a dedicated NLO QCD computation for the  interference  term and did not find a strong
indication 
that this might be the case.  Nevertheless, the moderate size of the NLO QCD corrections is the consequence of
a very strong cancellations between double
and single logarithms of the charm mass.  It is unclear if this cancellation
persists  in higher orders. Hence, resummation of ${\cal  O}(\ln\mc)$-enhanced  terms for this process is quite desirable.

\vspace*{1cm}
\hspace*{-0.7cm}{\bf Acknowledgments}\;
This research is partially  supported by the Deutsche Forschungsgemeinschaft
(DFG, German Research Foundation) under grant
396021762 - TRR 257.

\appendix

\section{Extraction of the ${\cal O}(\ln\mc)$-enhanced  contributions in the real
  corrections}
\label{sec:subt}
In this appendix we describe a procedure to extract ${\cal O}(\ln m_c)$ contributions to real emission corrections.
They arise because of the
quasi-singular behavior of real emission amplitudes in the soft or
collinear limits  involving  charm quarks. The potential singularities in these
limits are regulated by the charm mass leading
to an appearance of ${\cal O}(\ln m_c)$-enhanced  terms when integrated over relevant phase spaces.
To extract logarithms of $m_c$, we  subtract approximate expressions from exact matrix elements
that make the difference integrable in the $m_c \to 0$ limit and integrate the subtraction terms
over unresolved phase space to explicitly extract logarithms of $m_c$.

As an example, we consider the gluon-gluon partonic channel, i.e.
\begin{align}
  \label{eq:ggHccx-proc}
  g(p_1) + g(p_2)
  \longrightarrow
  H(p_H) + c(p_3) + \bar{c}(p_4)
  \,,
\end{align}
and discuss the extraction of ${\cal O}(\ln m_c  )$ terms in detail. This channel is suitable for such a discussion
since, if the charm-quark mass is kept
finite, it is free of soft and collinear divergences. Hence,
all relevant contributions can be computed numerically for small but finite $m_c$,
and used to validate  formulas where logarithms of $m_c$ have been extracted and $m_c \to 0$ limit has been
taken where appropriate.

As we already mentioned in the main text, we use
the nested soft-collinear subtraction scheme which,
at this order, is equivalent to the FKS subtraction
scheme~\cite{Frixione:1995ms,Frixione:1997np}. The details of the subtraction scheme 
can be found in the 
literature~\cite{Caola:2017dug,Caola:2019nzf,Caola:2019pfz,Asteriadis:2019dte}
and we do not repeat this discussion here.
Nevertheless, the treatment of quasi-collinear and quasi-soft
singularities related to the emission of {\it massive} charm quarks is new and requires an explanation.

We  focus on the interference contribution between the
Yukawa-like and the $\ggh$-like production mechanisms in the process
Eq.~\eqref{eq:ggHccx-proc}. The interference term is non-zero only if
helicity flip on the charm line occurs. Furthermore, the presence of
such a helicity flip causes the usual factorization formulas to break
down and the singular limits need to be explicitly analyzed.  We note
that, thanks to the symmetry of the squared amplitude for the process
in Eq.~\eqref{eq:ggHccx-proc} under the  exchange of $c$ and $\bar c$,  we can  consider only
the case where $\bar{c}$ quark
becomes  soft or quasi-collinear to one of the other
partons. The case when both $c$ and $\bar c$ become unresolved is impossible since
we require a charm jet in the final state.

The quasi-singular limits which appear in this channel are related to
the soft-quark limit $S_4$ with $E_4 \sim \mc$ and the three collinear
limits $C_{4i}$ with $i=1,2,3$ where $(\scr{4}{i})\sim\mc^2$.
Performing an iterative subtraction of these singular limits, we find
\begin{align}
  \begin{split}
    \label{eq:gg-fullcontrib}
    \langle \FLM(1_g,2_g; 3_c, 4_{\bar c}) \rangle
    ={}&
    \sum_{i=1}^{3}
    \langle (1-C_{4i}) (1-S_4) \omega^{(i)}_{123} \FLM(1_g,2_g; 3_c, 4_{\bar c}) \rangle
    \\
    &
    +\langle C_{4i} (1-S_4) \omega^{(i)}_{123} \FLM(1_g,2_g; 3_c, 4_{\bar c}) \rangle
    \\
    &
    +\langle S_4 \FLM(1_g,2_g; 3_c, 4_{\bar c}) \rangle
    \,,
  \end{split}
\end{align}
where the first term on the right-hand side denotes the
fully-regulated contribution and the second and third terms are the
collinear and the soft integrated subtraction terms. The factors
$\omega^{(i)}_{123}$ are the weights that describe  various collinear
sectors. They read
\begin{align}
  \omega^{(i)}_{123}
  ={}&
       \frac{1}{\rho_{4i}} \cdot
       \left[
        \frac{1}{\rho_{41}}
       +\frac{1}{\rho_{42}}
       +\frac{1}{\rho_{43}}
       \right]^{-1}
       \,,
       \label{eq.a3aa}
\end{align}
with $\rho_{4i} = 1 - \cos\theta_{4i}$. We note that, since all
$m_c \to 0$ singularities are subtracted in the fully-regulated term in
Eq.~\eqref{eq:gg-fullcontrib}, the $\mc\to0$ limit can be immediately
taken there. On the other hand, the integrated subtraction
terms in the second line of Eq.~\eqref{eq:gg-fullcontrib} require
care to capture all the ${\cal O}(\ln \mc ) $-terms and constants which survive
the $\mc\to0$ limit.

In the remaining part of this section, we discuss in detail the
integration of the subtraction terms. We first focus on the soft
subtraction term, i.e. the last term in Eq.~\eqref{eq:gg-fullcontrib}, and
then the integration of the collinear subtraction term, i.e. the first
term in the second line of Eq.~\eqref{eq:gg-fullcontrib}.

\subsection{Integration of the soft-quark subtraction terms}
\label{sec:gg-soft-int}
Consider the soft-quark subtraction term
\begin{align}
  \langle S_4 \FLM(1_g,2_g; 3_c, 4_{\bar c}) \rangle
  \,.
\end{align}
To compute it, we need to know the behavior of the amplitude in the  limit
$p_4 \sim m_c \to 0$  and then integrate it over the phase space of the charm anti-quark with momentum
$p_4$.

Although, normally, soft (gluon) emissions factorize into a product of an eikonal factor and a tree-level matrix element squared,
a similar formula for soft-quark emission,  relevant for helicity-flipping processes,  does not exist.
Hence,  we determine  the soft-quark limit of the interference    by
studying an explicit expression of  the
amplitude for the process in Eq.~\eqref{eq:ggHccx-proc} in the limit $p_4 \sim m_c \to 0$. We find
\begin{align}
  \begin{split}
    \label{eq:soft-limit-gg}
    S_4
    \interf\left[
      |\amp(1_g,2_g; 3_c, 4_{\bar{c}})|^2
    \right]
    \sim{}&
\frac{ (2\Cf-\Ca) (\scr{1}{2}) }{ (\scr{1}{4}) (\scr{2}{4}) }
     F_{12}(p_1,p_2,p_3) 
    \\
    &
    +  \frac{ \Ca (\scr{1}{3}) }{ (\scr{1}{4}) ( m_c^2 + \scr{3}{4} ) } \; F_{13}(p_1,p_2,p_3) 
    \\
    &
    +  \frac{ \Ca (\scr{2}{3}) }{ (\scr{2}{4}) ( m_c^2 + \scr{3}{4} ) }  F_{23}(p_1,p_2,p_3) \,,
  \end{split}
\end{align}
where functions $F_{ij}$ depend on the momenta $p_1$, $p_2$ and
$p_3$ {\it only}. We emphasize that these functions are different from the leading-order  interference contribution.
The massless limit, $\mc\to0$, can be now taken everywhere except
for the eikonal factors and  the phase-space measure of the
unresolved parton $[\d{p_4}]$.

We write the integrated soft subtraction term as follows
\begin{align}
  \begin{split}
  \langle S_4 \FLM(1_g,2_g; 3_c, 4_{\bar c}) \rangle
  ={}&
  (2\Cf-\Ca)
  \langle
  F_{12}(p_1,p_2,p_3) \cdot
  \int \frac{ [\d{p_4}] (\scr{1}{2}) }{ (\scr{1}{4}) (\scr{2}{4}) }
  \rangle
  \\
  &
  +\Ca
  \langle
  F_{13}(p_1,p_2,p_3)
  \cdot
  \int \frac{ [\d{p_4}] (\scr{1}{3}) }{ (\scr{1}{4}) (\mc^2+\scr{3}{4}) }
  \rangle
  \\
  &
  +\Ca
  \langle
  F_{23}(p_1,p_2,p_3)
  \cdot
  \int \frac{ [\d{p_4}] (\scr{2}{3}) }{ (\scr{2}{4}) (\mc^2+\scr{3}{4}) }
  \rangle
  \,,
  \end{split}
\end{align}
where  $\langle\ldots\rangle$ denotes the phase space integration
and the relevant soft integrals can be found in
Appendix~\ref{sec:soft-integrals}. We stress  that  soft
integrals are finite in four dimensions since they are naturally
regulated by the charm-quark mass $\mc$.

\subsection{Integration of the quasi-collinear subtraction terms}
\label{sec:gg-coll-int}

In this subsection, we  discuss how to define and compute the soft-subtracted quasi-collinear limits
of the interference contribution using the process in Eq.~(\ref{eq:ggHccx-proc}) as an example. We  focus on
the sector $43$ where $c$ and $\bar c$ become collinear to each other.
The relevant  quantity reads\footnote{We note that weight factors introduced in Eq.(\ref{eq.a3aa})
  do not appear in the collinear limits.}
\begin{align}
  \label{eq:int-coll-4i}
  \langle
  C_{43}(1-S_4) \FLM(1_g,2_g; 3_c, 4_{\bar c})
  \rangle\,.
\end{align}
To proceed further, we split the above
formula into collinear and soft-collinear terms
\begin{align}
  \langle C_{43}(1-S_4) \FLM(1_g,2_g; 3_c, 4_{\bar c}) \rangle
  ={}&
       \langle C_{43} \FLM(1_g,2_g; 3_c, 4_{\bar c}) \rangle
       - \langle C_{43} S_4 \FLM(1_g,2_g; 3_c, 4_{\bar c}) \rangle
       \,.
       \label{eq.a8}
\end{align}

We first   discuss the collinear subtraction term $\langle C_{43} \FLM(1_g,2_g; 3_c, 4_{\bar c}) \rangle$ defined
as follows
\begin{align}
  \begin{split}
    \label{eq:c43-flm1234-int}
    \langle C_{43} \FLM(1_g,2_g; 3_c, 4_{\bar c}) \rangle ={}& \sum_{i=1}^2
      \int [\d{p_H}][\d{p_3}][\d{p_4}] (2\pi)^4 \delta\left(p_{12}-p_H-p_3-{\bar p_4}\right)  \\
    & \hphantom{\sum_{i=1}^2 \langle} \times\frac{1}{(p_3 + p_4)^2}\frac{\scr{i}{3}}{p_{i}\!\cdot\!{\bar p_{4}}}{C}_{i}(1_g,2_g,3_c,\bar{4}_{\bar c})
    \,,
  \end{split}
\end{align}
In the above equation, the functions ${C}_{1,2}$ depend on the momenta
$p_1$, $p_2$, $p_3$ and $\bar{p}_4$. The bar over momentum $p_4$
indicates that the relevant collinear limit has been taken, i.e.
\begin{align}
  C_{43} p_4 = (E_4, \beta_4 \vec{n}_3) \equiv \bar{p}_4\,.
\label{eq.A9}
\end{align}
Note that in Eq.~(\ref{eq.A9})    $\beta_4 = \sqrt{1-\mc^2/E_4^2}$ is the velocity of $\bar c$ and $\vec{n}_3$  is
a unit vector pointing in the direction of momentum $\vec p_3$.
We note that in Eq.~(\ref{eq:c43-flm1234-int}) the massless limit  $\mc\to0$
has  not been taken.  We also note that  the functions $C_{1,2}$ are regular in the
soft-quark limit, $E_4 \to m_c \sim 0$.

Our goal is to  extract all ${\cal O}(\ln \mc) $ terms arising from
Eq.~\eqref{eq:c43-flm1234-int} and take the massless limit
after that.  To do so, we add and subtract the
soft limits of the functions $C_i$
\begin{align}
  \label{eq:c-split-c43}
  {C}_{i}(1_g,2_g,3_c,\bar{4}_{\bar c})
  ={}&
       \big[
       {C}_{i}(1_g,2_g,3_c,\bar{4}_{\bar c})
       -
       {C}_{i,\rm soft}(1_c,2_g,3_c)
       \big]
       +
       {C}_{i,\rm soft}(1_c,2_g,3_c)
       \,,
\end{align}
where ${C}_{i,\rm soft}(1_c,2_g,3_c) = {C}_{i}(1_g,2_g,3_c,0)$. The
above procedure splits the integral in Eq.~\eqref{eq:c43-flm1234-int}
into two parts: the \emph{regulated} integral that contains the
expression in the square bracket in Eq.~\eqref{eq:c-split-c43} and the
\emph{soft} part.
In the regulated part, the soft divergence  at $E_4 = 0$ has been regularized.
This implies that, after integrating $1/(p_3+p_4)^2$ over the relative angle between $p_3$ and $p_4$ and extracting
logarithms of $m_c$ from this angular integral, we can set  $m_c$ to zero  everywhere else right away. We obtain
\begin{align}
  \label{eq:coll-reg-integral}
  \langle
  C_{43}
  \FLM(1_g,2_g; 3_c, 4_{\bar c})
  \rangle_{\rm reg}
  ={}&
       \frac{1}{(2\pi)^2}
       \sum_{i=1}^2
       \int [\d{p_H}][\d{p_3}] (2\pi)^4 \delta\left(p_{12}-p_H-p_{34}\right)
       \nonumber\\
     &
       \int_0^1 \frac{z \, \d{z}}{1-z}
       \Big[
       C_{i}(1_g,2_g,z 34,(1-z) 34)
       -
       {C}_{i,\rm soft}(1_c,2_g,z 34)
       \Big]
       \nonumber\\
     &
       \times
       \Big[
       \ln(2E_{34}/\mc) + \ln(1-z) + \ln(z)
       \Big]
       \,,
\end{align}
where we have used the fact that in $\mc\to0$ limit we can write
$p_3 = zp_{34}$ and $\bar p_4 = (1-z)p_{34}$ for $p_{34}^2 = 0$.

We will now discuss the soft part of the collinear subtraction term. It reads
\begin{align}
  \begin{split}
    \label{eq:c43-flm1234-int-soft}
    \langle
    C_{43}
    \FLM(1_g,2_g; 3_c, 4_{\bar c})
    \rangle_{\rm soft}
    ={}&
    \sum_{i=1}^2
      \int [\d{p_H}][\d{p_3}][\d{p_4}]
      (2\pi)^4 \delta\left(p_{12}-p_H-p_3-{\bar p_4}\right)
    \\
    &
    \times\frac{1}{(p_3 + p_4)^2}
    \frac{\scr{i}{3}}{p_{i}\!\cdot\!{\bar p_{4}}}
    {C}_{i,\rm soft}(1_c,2_g,3_c)
    \,.
  \end{split}
\end{align}
We emphasize that this term still contains soft singularity and, for this reason,
the   $\mc\to0$ limit cannot be taken.  However, it is convenient
to combine this integral with the
soft-collinear subtraction term
$\langle C_{43} S_4 \FLM(1_g,2_g; 3_c, 4_{\bar c}) \rangle$, c.f.  Eq.~(\ref{eq.a8});
if this is done, the required computations simplify significantly.

The  soft-collinear integrated subtraction term in
sector 43 reads
\begin{align}
  \begin{split}
    \label{eq:coll-soft-integral}
    \langle
    C_{43} S_4
    \FLM(1_g,2_g; 3_c, 4_{\bar c})
    \rangle
    ={}&
    \sum_{i=1}^2
    \int [\d{p_H}][\d{p_3}][\d{p_4}]
    (2\pi)^4 \delta\left(p_{12}-p_H-p_3\right)
    \\
    &
    \hphantom{\sum_{i=1}^2}
    \times\frac{2 \Ca F_{i3}(p_1,p_2,p_3)}{(p_3 + p_4)^2}\frac{\scr{i}{3}}{p_{i}\!\cdot\!{\bar p_{4}}}
    \,.
  \end{split}
\end{align}
To derive this result we used the soft-limit of the
interference amplitude reported in Eq.~\eqref{eq:soft-limit-gg}. We
emphasize that, since the soft operator is present on the left hand side in the above
equation, the soft-quark momentum $p_4$ is removed  from the
energy-momentum conserving delta-function. Moreover, since
\begin{align}
  C_{i,\rm soft}(1_c,2_g,3_c)
  ={}&
       2 \Ca F_{i3}(p_1,p_2,p_3)
       \,,
\end{align}
the  two integrals in Eqs.~(\ref{eq:c43-flm1234-int-soft},\ref{eq:coll-soft-integral})
appear to be  the same up to
the argument of the delta-functions.
We combine the two integrals and find
\begin{align}
  \label{eq:coll-soft-combined}
  \langle
  C_{43}
  \FLM(1_g,2_g; 3_c,&  4_{\bar c})
                      \rangle_{\rm soft}
                      -
                      \langle
                      C_{43} S_4
                      \FLM(1_g,2_g; 3_c, 4_{\bar c})
                      \rangle
                      ={}
                      \nonumber\\
  ={}
                    &\sum_{i=1}^2
                      \int [\d{p_H}][\d{p_3}][\d{p_4}]
                      (2\pi)^4
                      \Big[
                      \delta\left(p_{12}-p_H-p_3-\bar{p}_4\right)
                      -
                      \delta\left(p_{12}-p_H-p_3\right)
                      \Big]
                      \nonumber\\
                    &
                      \hphantom{\sum_{i=1}^2}
                      \times\frac{2 \Ca F_{i3}(p_1,p_2,p_3)}{(p_3 + p_4)^2}\frac{\scr{i}{3}}{p_{i}\!\cdot\!{\bar p_{4}}}
                      \,.
\end{align}

To proceed further, we note that it is straightforward to integrate over directions of the quark with momentum
$p_4$ but integration over its energy is more involved. It is convenient to
split the $E_4$  integration into two regions by introducing an auxiliary parameter $\sigma$
\begin{align}
  1
  ={}&
       \Theta(E_4 - \sigma) + \Theta(\sigma - E_4)
       \,.
\end{align}
We choose $\sigma$ to satisfy the following inequality $\mc \ll \sigma \ll E_3$. For small energies,
$E_4 < \sigma \ll E_3$, we can drop the momentum $\bar{p}_4$ from the
energy momentum conserving delta-function which leads to
\begin{align}
  \begin{split}
    \Big[
    &\delta\left( p_{12}-p_H-p_3-\bar{p}_4 \right)\big[ \Theta(E_4 - \sigma) + \Theta(\sigma - E_4)   \big]
    -
    \delta\left(p_{12}-p_H-p_3\right)
    \Big]
    ={}
    \\
    &
    \hspace{1cm}
    ={}
    \Big[
    \delta\left( p_{12}-p_H-p_3-\bar{p}_4 \right)
    -
    \delta\left(p_{12}-p_H-p_3\right)
    \Big] \Theta(E_4 - \sigma)
    + \mathcal{O}( \sigma/E_3 )
    \,.
  \end{split}
\end{align}
This relation implies that the integrand in
Eq.~\eqref{eq:coll-soft-combined}  is non-vanishing only in the high-energy domain where 
$E_4 > \sigma \gg \mc$ and, therefore, the limit $\mc\to0$ can be taken.
This leads to the following expression
\begin{align}
  \label{eq:a18}
  \langle
  C_{43}
  \FLM(1_g,&2_g; 3_c, 4_{\bar c})
                      \rangle_{\rm soft}
                      -
                      \langle
                      C_{43} S_4
                      \FLM(1_g,2_g; 3_c, 4_{\bar c})
                      \rangle
                      ={}
                      \nonumber\\
                      = & \frac{C_A}{(2\pi)^2} \sum \limits_{i=1}^{2}
                      \int [{\rm d} p_H] [{\rm d} p_{34} ] (2\pi)^4 \delta(p_{12} - p_H - p_{34} )
                      \nonumber\\
&                       \Bigg \{
                      \int \limits_{z_{\rm min} }^{z_{\rm max}} \frac{{\rm d}z  }{1-z}
                       \ln \left( \frac{2E_{34}(1-z) z}{\mc} \right )
                       \Big[  z \theta(z) F_{i3} (p_1,p_2,z p_{34} )
                     -                                            F_{i3} (p_1,p_2,p_{34} ) \Big]
\nonumber\\
&   + \int \limits_{z_{\rm min} }^{z_{\rm max}}  \frac{ {\rm d}z  }{1-z}
                       \ln \big( (2-z) z \big)
                             F_{i3} (p_1,p_2,p_{34} )
                      \Bigg \}.
\end{align}
To arrive at Eq.~(\ref{eq:a18})  we introduced the four-momentum
$p_{34} = p_3 + \bar{p}_4$ and  a variable $z$ such that $p_3 = z p_{34}$ in terms that contain  the delta-function
$\delta{(p_{12}-p_H-p_3-\bar{p}_4)}$.  In 
terms that contain the delta-function  $\delta{(p_{12}-p_H-p_3)}$,  we set $(1-z)=E_{4}/E_{3}$,
rename  $p_{3}$ into $p_{34}$ and set $\sigma \to 0$. The lower integration boundary $z_{\rm min}$ is given
by $z_{\rm min} = 1 - \Emax/E_{34} < 0$.

In total, the integrated collinear term
$\langle C_{43}(1-S_4) \FLM(1_g,2_g; 3_c, 4_{\bar c}) \rangle$ is
given by a sum of expressions in Eqs.~(\ref{eq:coll-reg-integral},\ref{eq:a18}).
We describe  a numerical
check of validity of this result in the following section.

\subsection{Numerical checks}
Since the cross section of the gluon-gluon  channel, Eq.~\eqref{eq:ggHccx-proc},
is  finite
as long as we keep the non-zero charm mass, analytical results
derived in the previous section can be checked numerically by computing $\sigma_{gg \to Hc \bar c}$
explicitly for small  values
of the charm mass without any approximation.

\begin{figure}[h!]
  \centering
  \includegraphics[width=0.8\textwidth]{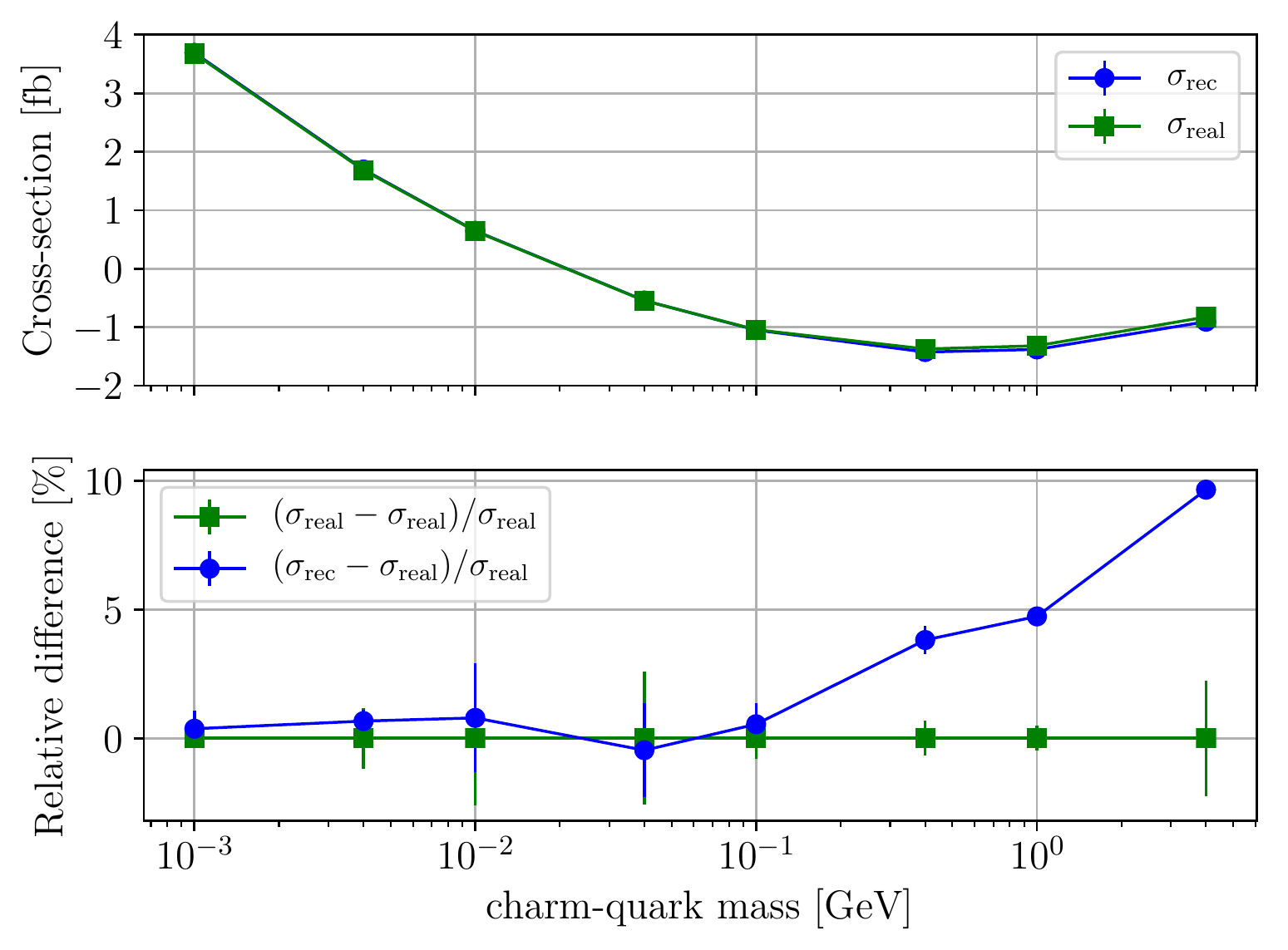}
  \caption{The cross section of the $gg\to H\cc$ process calculated by
    a direct integration of the matrix element with non-zero
    charm-quark mass, $\sigma_{\rm real}$ (green rectangles), and
    reconstructed using procedure described in previous subsections,
    $\sigma_{\rm rec}$ (blue circles). We employ the same parameters and kinematic constraints
    as in the main text.}
  \label{fig:plots-coll}
\end{figure}

The comparison is shown in Figure~\ref{fig:plots-coll}. We use fiducial cuts described
in the main text and  compare hadronic  contributions  to  the interference for $gg$ partonic channel
computed in two
different ways. Green points (rectangles) show the results of the computation without any approximation, i.e.
by directly  integrating   the matrix element squared.
Blue points (circles) show the results of the computation that relies on the expansion around $m_c \to 0$ limit, as described in  previous subsections.
The two results should agree for small values of $m_c$. The upper panel of
Figure~\ref{fig:plots-coll} shows the absolute values of the
interference cross section in the $gg$ partonic channel obtained with
the two methods, while their  difference
is shown in the lower panel. We see a better and better agreement
between the two results  as we  mover to smaller and smaller values of the charm-quark mass.
This indicates that the $\mc$-dependence of the interference contribution
is properly reconstructed.

\section{Soft-quark integrals}
\label{sec:soft-integrals}
In this section we list integrals that are required for the
integrated soft-quark subtraction terms.
We need a number of integrals depending on the type and configuration
of the emitters $p_a$ and $p_b$ as well as the propagator appearing in
the eikonal factor.

We note that we are interested only in the terms that contain
logarithms of the charm-quark mass and constant terms, but we drop all
power-suppressed terms which vanish in the $\mc\to0$ limit.
All integrals are computed in $d=4$ dimensions since all singularities
are naturally regulated by the charm-quark mass.

The phase-space measure for a massive-quark emission, $p_4^2 = \mc^2$,
reads
\begin{align}
  [\d{p_4}]
  ={}&
  \frac{k_4^2 \d{k_4}}{2 E_4}
  \frac{ \d\Omega^{(3)} }{(2\pi)^3}
  \Theta( \Emax - E_4)
  \Theta( E_4 - \mc )
  \,,
\end{align}
where $k_4$ is the length of $\vec{p}_4$ momentum, $\d\Omega^{(3)}$ denotes angular integration and $\Emax$ is the
usual energy cutoff of the nested soft-collinear subtraction scheme.
In the remaining part of this section, we list soft-quark
integrals that are needed to obtain integrated soft-quark subtraction
terms, see Section~\ref{sec:gg-soft-int} for details.\footnote{Similar
expressions to those in Section~\ref{sec:gg-soft-int} can be derived
for other partonic channels featuring soft-quark singularities,
i.e. $cc$ and $\cc$.}

\paragraph{Two massless emitters:}
Two emitters $a,b$  have  four-momenta $p_a=E_a(1,\vec{n}_a)$ and
$p_b = E_b(1,\vec{n}_b)$, respectively. Both four-momenta are light-like $p_a^2 = p_b^2 = 0$.  Vectors $\vec{n}_a$
and $\vec{n}_b$ describe direction of flight of the emitters; we refer to  the  opening
angle between $\vec n_a$ and $\vec n_b$ as $\theta_{ab}$.

The soft integral reads
\begin{align}
  \int
  \frac{ [\d{p_4}] (\scr{a}{b}) }{ (\scr{a}{4})(\scr{b}{4}) }
  ={}&
  \frac{1}{(2\pi)^2}
  \left[
      \ln^2(2 s_{ab} \Emax/\mc)
    - \frac{\pi^2}{12}
    + \frac{1}{2}\LiTwo(c_{ab}^2)
    \right]
  \,,
\end{align}
where we used  $s_{ab} = \sin(\theta_{ab}/2)$ and $c_{ab} =
\cos(\theta_{ab}/2)$.

\paragraph{One massive and one massless emitters:}
Two emitters $a,b$ have four-momenta $p_a=E_a(1,\vec{n}_a)$ and
$p_b = E_b(1,\beta_b\vec{n}_b)$, respectively. They satisfy  $p_a^2 = 0$ and $p_b^2 = \mc^2$.
We refer to  the  opening
angle between $\vec n_a$ and $\vec n_b$ as $\theta_{ab}$.
We require three soft integrals of this type
\begin{align}
  \begin{split}
    &  \int
    \frac{ [\d{p_4}] (\scr{a}{b}) }{ (\scr{a}{4})(\mc^2 + \scr{b}{4}) }
    ={}
    \frac{1}{(2\pi)^2}
    \bigg[
    \ln^2( 2 s_{ab} \Emax / \mc )
    - \frac{\pi^2}{12}
    \\
    &
    \hphantom{\frac{1}{(2\pi)^2}\bigg[}
    + \LiTwo(-\Emax/E_b)
    + \frac{1}{2} \LiTwo(c_{ab}^2)
    \bigg]
    \,,
    \\
       &  \int
    \frac{ [\d{p_4}] (\scr{a}{b}) }{ (\scr{a}{4})(\scr{b}{4}) }
    =
    \frac{1}{(2\pi)^2}
    \bigg[
    \ln^2(2 s_{ab} \Emax/\mc)
    + \frac{1}{4}\LiTwo(-\Emax^2/E_b^2)
    \\
    &
    \hphantom{\frac{1}{(2\pi)^2}\bigg[}
    - \frac{\pi^2}{12}
    +\frac{1}{2} \LiTwo(c_{ab}^2)
    \bigg]\, ,
    \\
    &     \int
    \frac{ [\d{p_4}] (\scr{a}{b}) }{ (\scr{a}{4})(\mc^2 - \scr{b}{4}) }
    =
    \frac{1}{(2\pi)^2}
    \bigg[
    -\ln^2(2 s_{ab} \Emax/\mc)
    + \LiTwo(1-\Emax/E_b)
    \\
    &
    \hphantom{\frac{1}{(2\pi)^2}\bigg[}
    -\frac{1}{2} \LiTwo(c_{ab}^2)
    + \ln(\Emax/E_b) \ln(\Emax/E_b - 1)
    - \frac{\pi^2}{12}
    \bigg]
    \,,
  \end{split}
\end{align}
where we used  $s_{ab} = \sin(\theta_{ab}/2)$ and $c_{ab} = \cos(\theta_{ab}/2)$.

\paragraph{Two massive emitters:}
Two emitters $a,b$ have four-momenta
$p_a=E_a(1,\beta_a\vec{n}_a)$ and $p_b = E_b(1,\beta_b\vec{n}_b)$, respectively.
They satisfy  $p_a^2 = p_b^2 = \mc^2$.
We refer to  the  opening
angle between $\vec n_a$ and $\vec n_b$ as $\theta_{ab}$.
In this case, we use  $\Emax =
E_a$.  We find 
\begin{align}
  \begin{split}
    & \int
    \frac{ [\d{p_4}] (\scr{a}{b}) }{ (\mc^2 - \scr{a}{4})(\mc^2 + \scr{b}{4}) }
    ={}
    \frac{1}{(2\pi)^2}
    \bigg[
    -\ln^2( 2s_{ab}\Emax/\mc)
    -\frac{\pi^2}{12}
    \\
    &
    \hphantom{\frac{1}{(2\pi)^2}\bigg[}
    -\frac{1}{2}\LiTwo(c_{ab}^2)
    -\LiTwo(-\Emax/E_b)
    \bigg]
    \,,
    \\
     &  \int
  \frac{ [\d{p_4}] (\scr{a}{b}) }{ (\mc^2 - \scr{a}{4})(\mc^2 - \scr{b}{4}) }
  =
  \frac{1}{(2\pi)^2}
  \bigg[
    \ln^2(2\Emax/\mc)
    + \frac{\pi^2}{4}
    \bigg]\, ,
  \end{split}
\end{align}
where we used $s_{ab} = \sin(\theta_{ab}/2)$ and $c_{ab} = \cos(\theta_{ab}/2)$.


\end{document}